# Towards a Metal–Organic Framework with Pore-Confined Electrons

Julia H. Baratta,[a] Andrew S. Rosen[a,*]

[a]Department of Chemical and Biological Engineering, Princeton University, Princeton, NJ 08544, United States

[*]Corresponding author: asrosen@princeton.edu

## Abstract

Electrides are an unconventional class of materials in which electrons are localized in crystallographic void spaces rather than solely around atomic nuclei, giving rise to appealing properties such as low work functions, strong electron-donating character, and even superconductivity. Here, we use *ab initio* methods to investigate metal–organic framework (MOF) electrides, a new class of materials that combines the interstitial electrons of electrides with the permanent porosity and chemical tunability of MOFs. These materials host pore-confined electrons: occupied electronic states localized in the pore space and with bands slightly below or crossing through the Fermi level. Using density functional theory calculations, we establish several design rules for stabilizing pore-confined electrons in MOFs via an anion–electron exchange process and identify candidate MOF electrides. As a proof-of-concept, we also demonstrate that the pore-confined electrons can directly facilitate chemical reactions, substantially lowering the activation barrier for $H_2$ dissociation without requiring adsorption at a surface site. We envision that pore-confined electrons in nanoporous materials may enable a fundamentally new type of catalysis in which chemical reactions take place in the pore space, driven by electron-centered active sites.

## Introduction

In conventional solid-state materials, electrons are either localized around atomic nuclei or shared between them in the form of chemical bonds. However, in a relatively rare class of materials known as electrides, some electrons are not associated with atomic nuclei at all. Instead, these electrons are localized in crystallographic void spaces analogous to charge-balancing anions.[1] These electrons are typically mobile and energetically accessible, leading to attractive material properties such as high electrical conductivity, low work functions, and strong electron-donating character.[1,2] To date, many different classes of electrides have been discovered, including organic electrides, inorganic electrides, and van der Waals layered electrides.[1–6]

Organic electrides, which were the focus of early research in this field, typically consist of alkali metal cations embedded in an organic molecular crystal, where the structure stabilizes the electrons released by the alkali metals.[5,6] While these materials display many of the interesting electronic properties associated with free electrons, they are often limited by thermal stability, although several room-temperature stable organic electrides have since been synthesized.[7–10] To combat these stability limitations, attention shifted towards designing inorganic materials with cavity-trapped electrons.

The most widely studied inorganic electride is derived from mayenite, a naturally occurring mineral in which charge-balancing anions (e.g. $O^{2-}$, $Cl^{-}$) can be exchanged with electrons to form the electride phase $[Ca_{12}Al_{14}O_{32}]^{2+}[2e^{-}]$.[11,12] These "non-nuclear" electrons are confined within sub-nanometer scale cages, and the resulting material is stable in air at temperatures up to 400 °C.[11] The electron-donating character of the mayenite electride has made it an attractive catalyst support for ammonia synthesis by facilitating the dissociation of the strong triple bond of $N_2$.[13–15] Although the cage-confined electrons of the mayenite electride participate only indirectly via charge transfer to supported metal nanoparticles, one can envision a porous electride with directly accessible electrons that could enable unique—and potentially enhanced—catalytic behavior.

Motivated by the cage-like structure of mayenite, we turn to metal–organic frameworks (MOFs), which are nanoporous solids constructed from metal nodes connected by organic linkers.[16] In addition to their high surface areas and permanent porosity, MOFs offer exceptional chemical tunability due to the vast design space of metal nodes and organic linkers, making them an ideal platform for investigating unique electronic structure properties.[17–19] We hypothesize that an appropriately designed MOF could exhibit an electronic ground state where the pore space stabilizes lone electrons. If realized, these pore-confined electrons could enable chemical reactions to be carried out directly within the electron-rich void space (Figure 1).

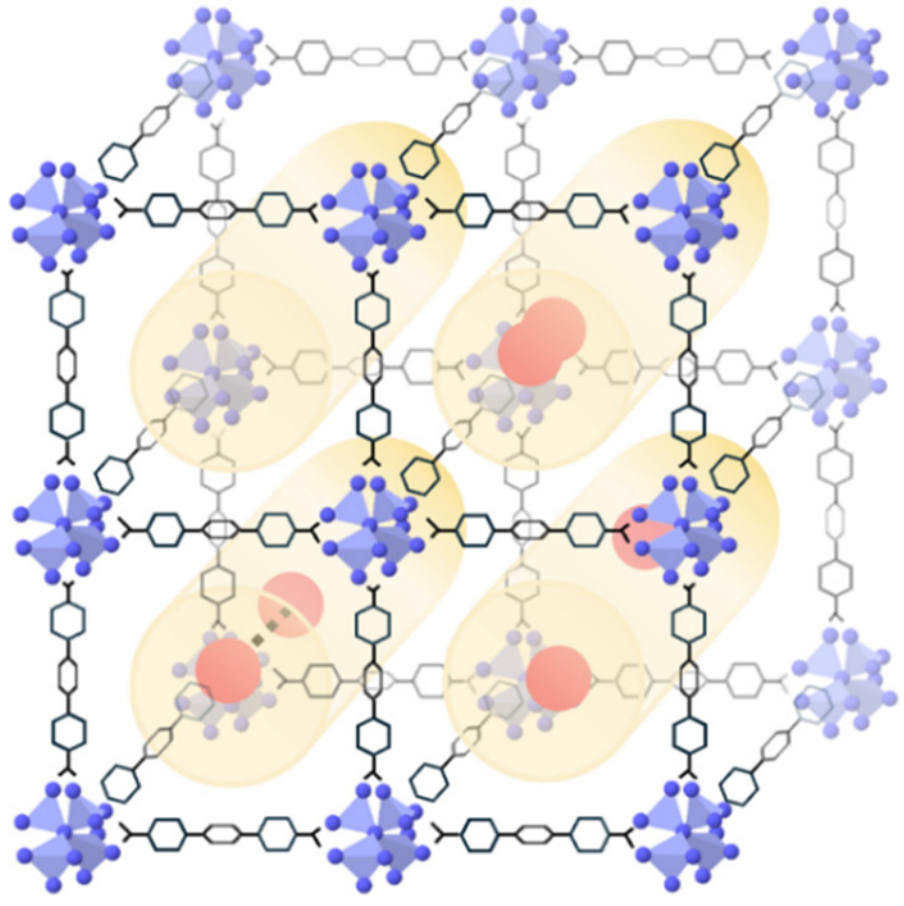

**Figure 1**. Depiction of a MOF with pore-confined electrons (yellow). A diatomic molecule (red) is shown dissociating upon interacting with the pore-confined electrons.

Recently, several studies have suggested that the pores of MOFs can support unusual electronic states. Goesten and Schoop,[20] for example, used group theory to postulate that the hypothetical model system $Pt_3(hib)_2$ (hib = hexaiminobenzene) would exhibit a pore band if synthesized. The pore band in $Pt_3(hib)_2$ is predicted to reside within the conduction band rather than the valence band, such that it is unoccupied in the ground state. As such, $Pt_3(hib)_2$ would not be classified as a typical electride,[20] and the electronic state is best thought of as analogous to the unoccupied interlayer band of graphite.[21] The pore conduction band can, however, become occupied when the MOF is doped with excess electrons (Figure S1b). In recent follow-up work by Berg and Goesten, the isostructural $Ni_3(hib)_2$ analogue was predicted to have a pore band residing only ~0.1 eV above the Fermi level and could, theoretically, become occupied at elevated pressures (e.g. 14 GPa),[22] although this may be beyond the stability limits of most 2D MOFs.[19] In separate work, Golomb and coworkers reported that when $Fe_2(Cl_2dhbq)_3$ (dhbq = 2,5-dihydroxybenzoquinone) is loaded with $[Me_2NH_2]^+$ (Me = methyl) cations beyond the experimentally suggested stoichiometry, a localized electronic state can reside between the molecular cations on the basis of density functional theory (DFT).[23] That said, this electronic state is not intrinsic to the MOF itself, and the ground-state configuration of the cations under realistic conditions has not been investigated. As such, there remain many open questions about the conditions that may stabilize occupied electronic states in the pore space of MOFs.

*Ab initio* methods based on DFT have proven to be particularly promising for characterizing the electronic structure of both known and proposed electrides.[24–28] Notably, data-driven approaches enabled by high-throughput DFT calculations have led to the identification of many plausible electride candidates in the literature.[29–37] Several inorganic materials first identified as electrides through *ab initio* calculations have since been experimentally synthesized, including $[Sc_2C][2e^-]$,[38,39] $[Y_2C][2e^-]$,[31,40] $[Sr_3CrN_3][e^-]$,[33,41] and the high-pressure electrides $[Na_2He][2e^-]$[42] and electron-neutral LiAlGe.[43] Despite this precedent, similar data-driven approaches have not previously been applied to identify and characterize MOF electrides.

In this work, we use high-throughput DFT calculations to identify MOFs with pore-confined electrons (PCEs), demonstrating the plausibility of MOF electrides. From the MOF electride candidates identified in our screening process, we characterize their electronic structure as well as their thermodynamic stability and identify several design principles. As expected based on prior literature, many of the most promising MOF electrides are composed of electropositive metals with rigid oxidation states.[44] In addition, we predict that post-synthetic anion–electron exchange of fluoride anions is likely the most thermodynamically favorable route for MOF electride activation. We conclude by focusing on a representative MOF electride candidate to demonstrate a new avenue for chemical reactivity in which reactions take place in electron-rich void space, facilitated directly by charge transfer from the PCEs.

## Results and Discussion

### *High-Throughput Screening to Identify MOF Electride Candidates*

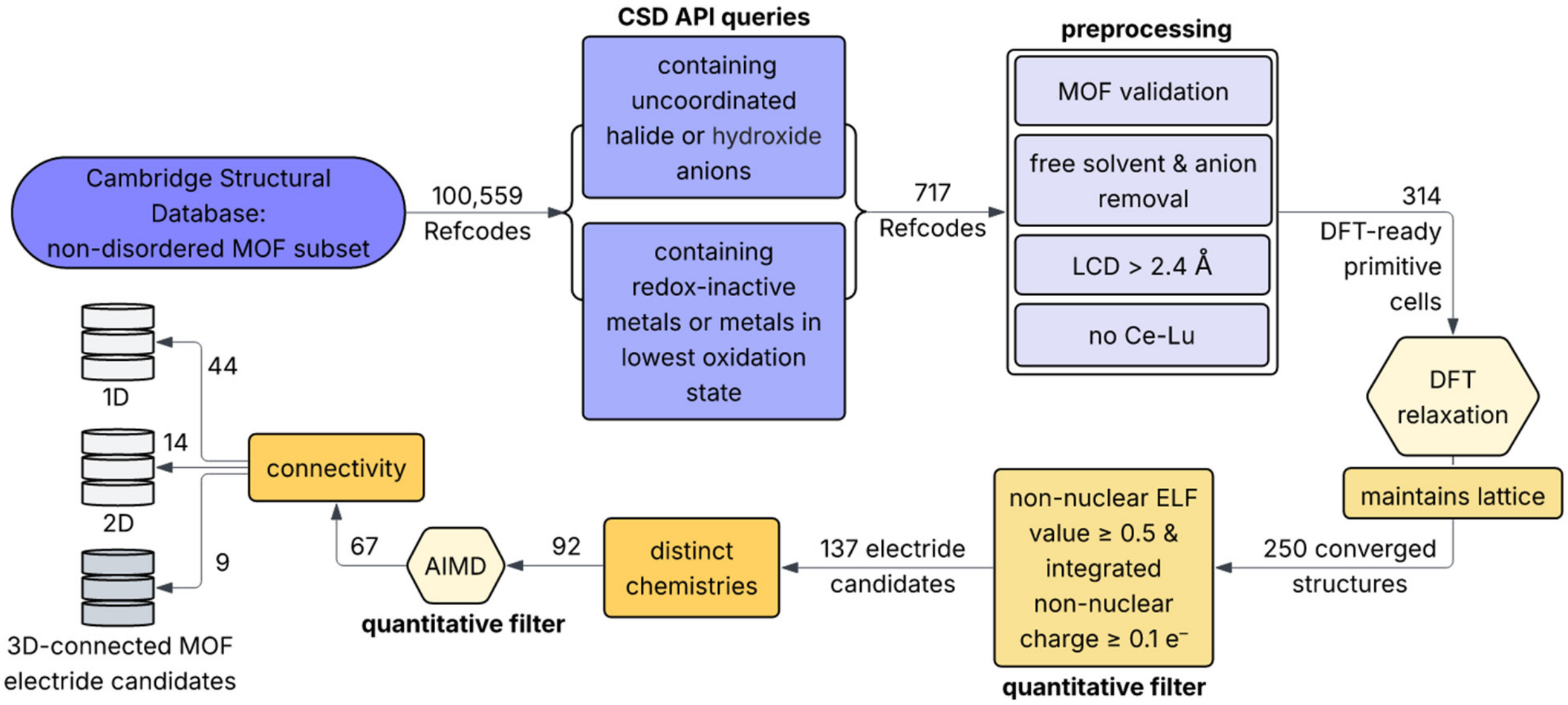

**Figure 2**. Workflow for the high-throughput screening process to identify MOF electride candidates. This workflow begins by querying the Cambridge Structural Database for MOFs that have charge-balancing anions and are likely to have redox-inactive metal centers. MOF structures were preprocessed and relaxed using DFT to yield the predicted electronic structure data. These MOF electride candidates were then screened for electride character both before and after brief *ab initio* molecular dynamics (AIMD) simulations.

Motivated by the precedent set by mayenite and other activated electrides,[14] we carried out a computational screening procedure (Figure 2) to identify MOFs that can stabilize pore-confined electrons following an anion–electron exchange process. We filtered the non-disordered MOF subset of the Cambridge Structural Database (CSD)[45] for MOFs with uncoordinated, charge-balancing anions and excluded structures with redox-active metals that were not in their lowest oxidation state. As described further in the Supplementary Information section, a suite of validation and preprocessing checks was carried out, leaving 314 MOFs for subsequent DFT calculations. To mimic the anion–electron exchange process used to synthesize activated electrides,[1,46] we removed the charge-balancing anions and relaxed each structure at the PBE-D3(BJ) level of theory.[47–50] This approach, which is analogous to prior computational studies of activated electrides,[51,52] converges to the electron density that minimizes the energy of the system—which may or may not be an electride state depending on where the excess electron density localizes. From this procedure, 250 MOFs were successfully optimized and considered for further electronic structure analyses.

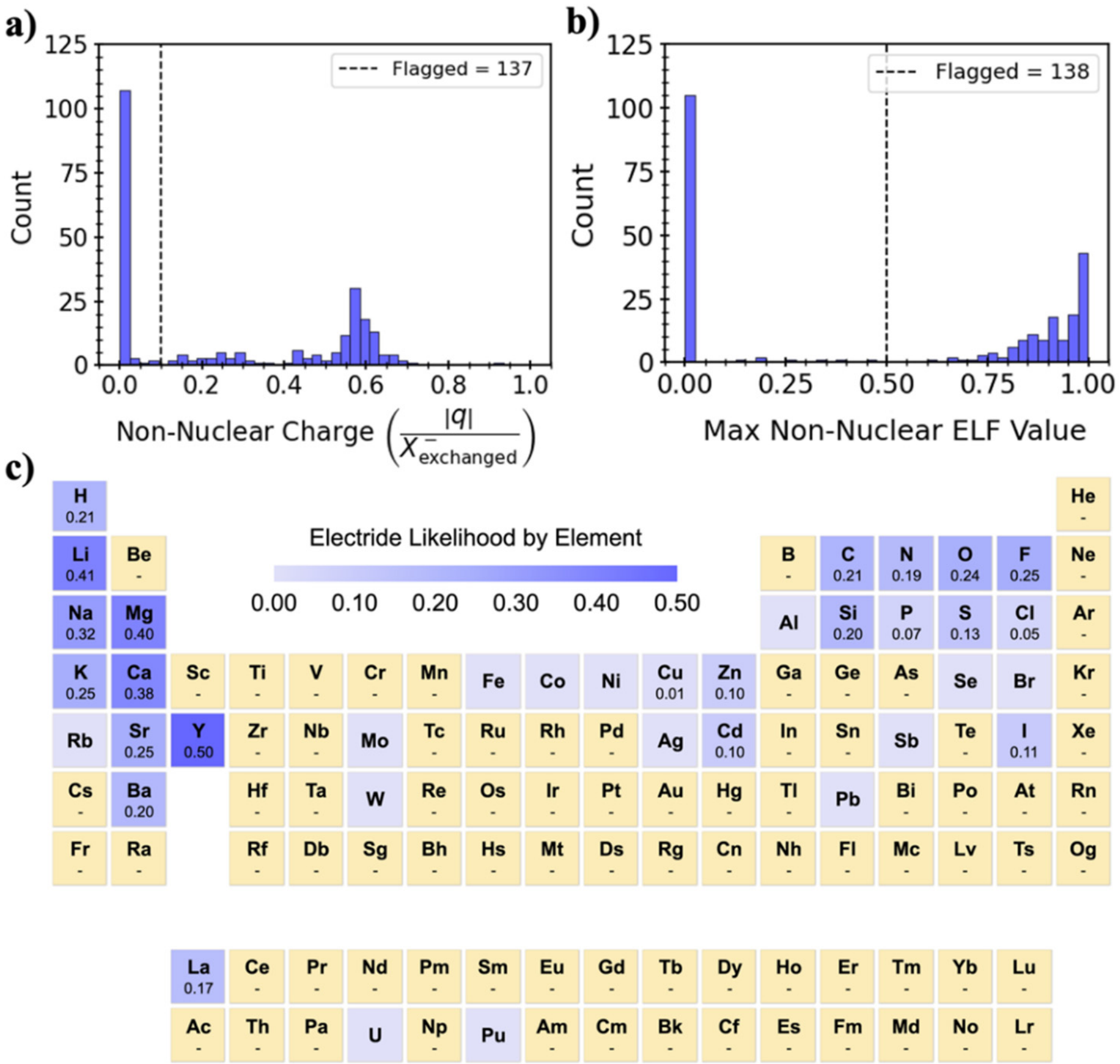

**Figure 3.** (a) Non-nuclear charge for each MOF in the 250 converged-structure subset, normalized on a per-anion-exchanged basis. (b) Maximum non-nuclear ELF value for each MOF in the 250 converged-structure subset. (c) The conditional probability of finding a MOF containing a given element in the 67-MOF subset with respect to the number of MOFs containing that element in the 314 DFT-ready subset. For example, of the 67 MOF electride candidates, all of them contain carbon, making the elemental conditional probability 67/314 = 0.21 for carbon. Yellow elements were not contained in the original dataset of 314 DFT-ready MOFs.

The degree of electride character can be evaluated based on the Electron Localization Function (ELF), which is a scalar field containing values ranging from zero (i.e. no electron localization) to one (i.e. highly localized electrons), with 0.5 normalized to that of a homogeneous electron gas.[53] Based on previous computational studies of electrides,[27,44] the MOFs with the greatest degree of electride character are expected to have high ELF values and substantial charge within their void space.

For many of the 250 DFT-optimized MOFs, the excess electron density delocalizes throughout the structure (e.g. around the linkers and/or nodes), signifying a lack of electride character. This is made apparent by the large fraction of MOFs with near-zero non-nuclear charge (Figure 3a) and non-nuclear ELF values (Figure 3b). To flag MOFs that may host pore-confined electrons, we adopted a maximum non-nuclear ELF criterion of $\geq$0.5 and a non-nuclear charge criterion of $\geq$0.1 $e^-$ per anion-exchanged. While these criteria do not by themselves guarantee electride character, they are useful baselines for the minimum consideration of electride character. As a point of reference, previous studies of known organic electrides predict non-nuclear charge values in the range 0.13–0.36 $e^-$.[26] We note here that the PBE-D3(BJ) level of theory is expected to underpredict non-nuclear ELF and charge values due to electron self-interaction error,[54,55] as we later show in Table 1. In total, 137 MOFs satisfy both criteria, of which 92 are retained after de-duplication.

To further probe the results of the MOF electride screening process, we evaluated the likelihood of finding a MOF electride given that it contains a particular element (Figure 3c). Based on our screening process, MOFs containing alkali metals or alkaline-earth metals have a high probability of adopting electride character following activation, which aligns with previous work on electrides.[27,44] Conversely, of the few transition metals included in this study, almost all of them did not exhibit an electride state, supporting our original hypothesis that MOF nodes containing redox-active metals would be less likely to stabilize free electrons.

As a final step in the screening process, we ran brief *ab initio* molecular dynamics (AIMD) simulations in the *NVT* ensemble at 300 K for each of the 92 distinct MOF electride candidates. In several cases, the pore-confined electrons abstract a labile hydrogen atom from a nearby linker. For instance, $[Al_2(H_4pmp)_3][6e^-]$ ($H_4pmp$ = N,N′-piperazinebis(methylenephosphonic acid); Refcode: KIMMIG) has zwitterionic linkers that readily become deprotonated to ultimately produce $H_2$ gas. In other cases, the finite-temperature dynamics of the framework structure quenched the electride state. As defined based on the same ELF and charge density criteria described above, a total of 67 candidates retained an electride state following the AIMD procedure. Of these, nine are classified by the CSD as being 3D-connected frameworks; the lower-dimensionality structures can generally be described as coordination polymers.

*Electronic Structure Properties of Select MOF Electride Candidates*

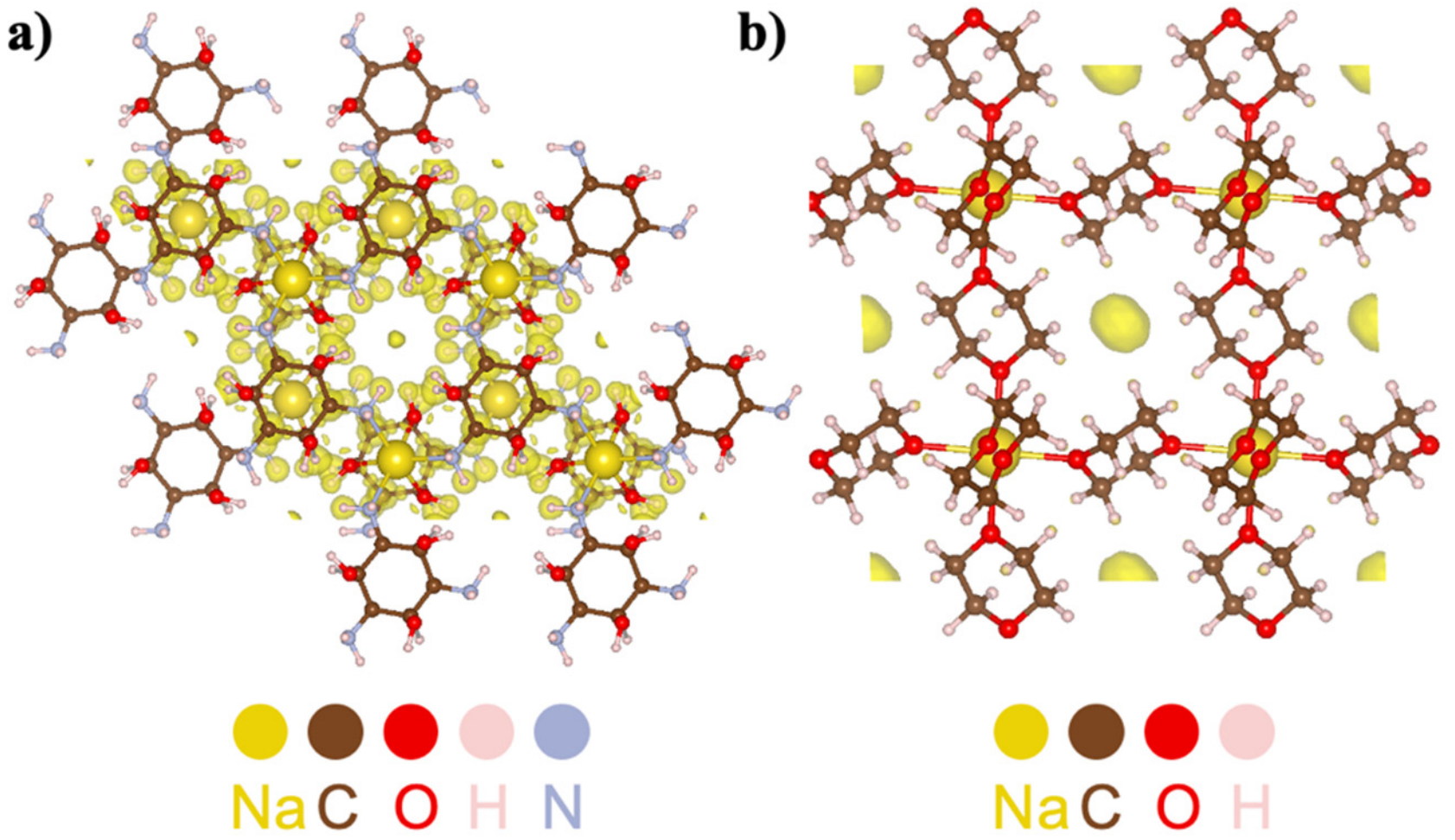

**Figure 4.** ELF visualizations at the HSE06 level of theory for two MOFs with significant electride character. The electride state is found in the pore space of each structure. (a) $[Na(taci)][e^-]$ (ELF isosurface = 0.900, spin-up channel, CSD Refcode: MEWDID). (b) $[Na(1,4\text{-dioxane})_3][e^-]$ (ELF isosurface = 0.999, spin-down channel, CSD Refcode: FITZOX).

For the sake of illustration, we highlight two representative examples of 3D-connected MOF electride candidates that emerged from the screening procedure: $[Na(taci)][e^-]$ (taci = 1,3,5-triamino-1,3,5-trideoxy-*cis*-inositol) and $[Na(1,4\text{-dioxane})_3][e^-]$, which are shown in Figures 4a and 4b, respectively. For both MOFs, localized electron density can be seen in the pore channels, as indicated by high ELF values (Figure 4, Figures S5–S6) and substantial charge in the void spaces (Table 1). These results hold for both the PBE-D3(BJ) level of theory and the HSE06 level of theory,[49,56,57] the latter of which predicts a greater degree of electron localization due to reduced electron self-interaction error (Table 1). As a point of comparison, the predicted non-nuclear charge for both MOF electride candidates is greater than that of the prototypical mayenite electride, which we predict to be 0.48 $e^-$ at the PBE-D3(BJ) level of theory when normalized by 2× the number of $O^{2-}$ anions exchanged. This fact—combined with the near-unity non-nuclear ELF values and relatively large distance between the pore-confined electrons and the nearby atoms—suggests that such materials would have substantial free-electron-like character if synthesized. For reference, the ELF plots and tabulated electronic structure data for all nine 3D-connected MOFs at the PBE-D3(BJ) level of theory are shown in Figure S7 and Table S1, respectively.

**Table 1**. Maximum non-nuclear ELF value and non-nuclear charge for the two highlighted MOF electride candidates. The maximum ELF value and non-nuclear charge are reported based on consideration of both the spin-up and spin-down channels. The non-nuclear charge is normalized on a per-anion-exchanged basis.

| MOF Electride Candidate | Parent Refcode | Non-Nuclear ELF Max | | Non-Nuclear Charge | |
|---|---|---|---|---|---|
| | | PBE-D3(BJ) | HSE06 | PBE-D3(BJ) | HSE06 |
| [Na(taci)][e⁻] | MEWDID[58] | 0.890 | 0.950 | –0.567 | –0.633 |
| [Na(1,4-dioxane)$_3$][e⁻] | FITZOX[59] | 0.999 | 1.000 | –0.686 | –0.755 |

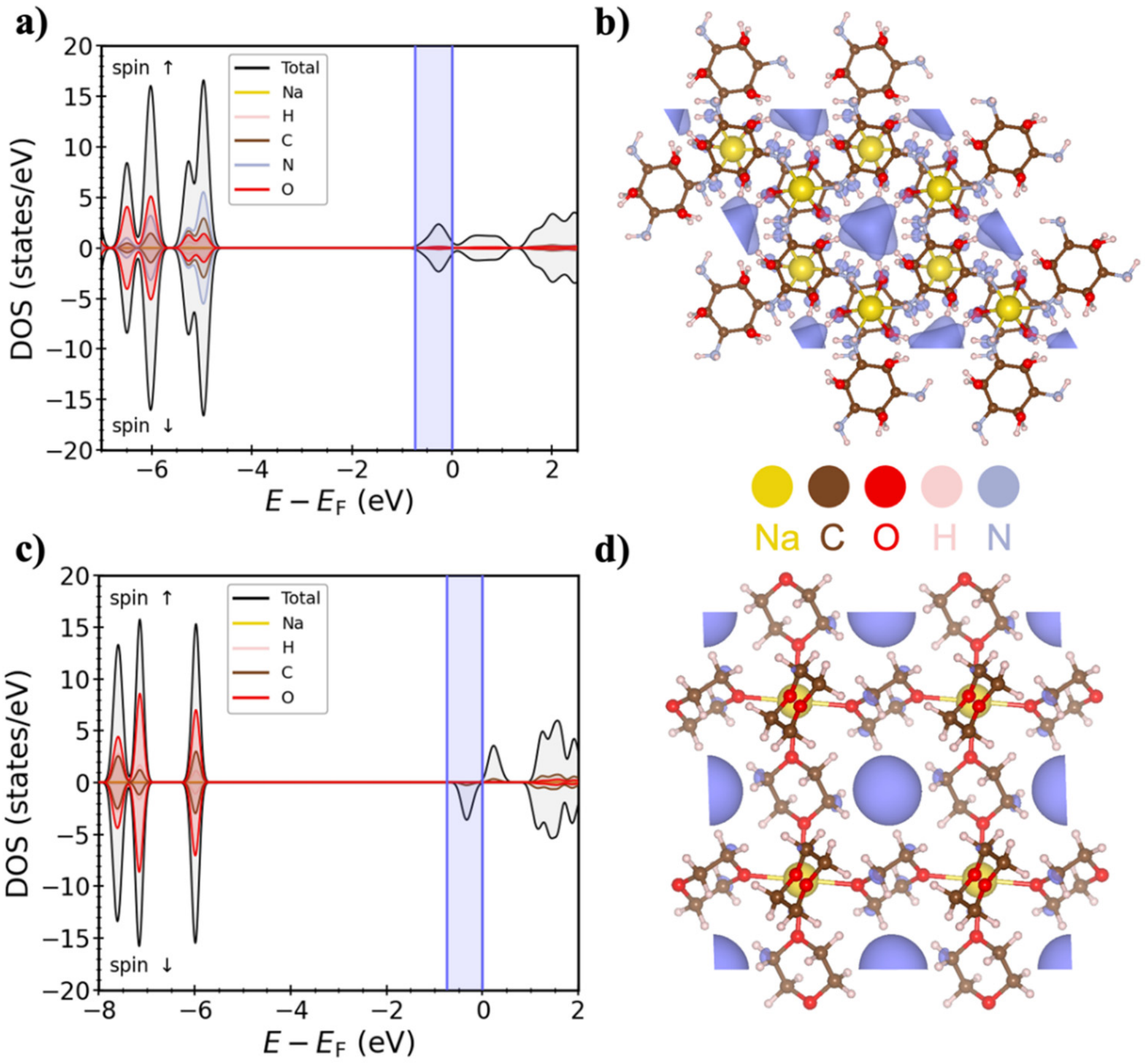

**Figure 5.** (a) Element-projected (colors) and total (black) electronic density of states (DOS) for [Na(taci)][e⁻]. The blue shaded region defined by –0.75 < $E - E_F$ < 0 eV corresponds to the band-decomposed charge density visualized in panel b. (b) Band-decomposed charge density for the electride state of [Na(taci)][e⁻] (isosurface = 0.0025 e⁻/bohr$^3$). (c) Element-projected (colors) and total (black) electronic density of states (DOS) for [Na(1,4-dioxane)$_3$][e⁻]. The blue shaded region defined by –0.75 < $E - E_F$ < 0 eV corresponds to the band-decomposed charge density visualized in panel d. A negative sign is used for the DOS of the spin-down channel for visual clarity in both DOS plots. (d) Band-decomposed charge density for the electride state of [Na(1,4-dioxane)$_3$][e⁻] (isosurface = 0.0016 e⁻/bohr$^3$). All data is at the HSE06 level of theory.

The predicted electronic DOS at the HSE06 level of theory for [Na(taci)][e⁻] has a well-defined, occupied electronic state near the Fermi level that is common among electrides (Figure 5a). Interestingly, the pore-confined electron density corresponding to [Na(1,4-dioxane)$_3$][e⁻] exists below the Fermi level of the spin-down channel only (Figure 5c). These asymmetric spin channels arise because the pore-confined electrons are themselves magnetic. The magnetic configuration shown in Figures 5c and 5d is associated with a net magnetic moment of –1.00 $\mu_B$ per cell, which is 0.15 eV more stable than a non-magnetic configuration at the HSE06 level of theory; that said, we note that the ground-state magnetic ordering of the pore-confined electrons (e.g. ferromagnetic vs. antiferromagnetic coupling) may be more complex. The presence of magnetic moments on the anionic electron at the HSE06 level of theory renders [Na(1,4-dioxane)$_3$][e$^-$] a magnetic electride—a subclass of electrides with particular relevance to the field of spintronics.[29,37,60] The band-decomposed charge densities for the occupied electronic states near the Fermi level clearly indicate that the charge density is localized within the pore space of both MOFs (Figures 5b and 5d). The DFT-predicted band structures further reveal the energy and dispersion of these bands relative to the Fermi level (Figure S8).

While the two MOF electride candidates mentioned in this section have both passed the AIMD filter described in Figure 2, we note that unbiased AIMD simulations will only exclude structures with particularly low deprotonation barriers. Since [Na(taci)][e$^-$] has –OH and –$NH_2$ groups that could, in principle, become deprotonated due to the electride state, we ran a transition state search to determine the deprotonation barrier. As shown in Figure S11, there is a predicted activation barrier of 0.95 eV that would likely prevent deprotonation from occurring except at elevated temperatures. We also considered a similar scenario for [Na(1,4-dioxane)$_3$][e$^-$]; however, cleavage of a hydrogen atom from a –$CH_2$ group of the dioxane linker is not predicted to be feasible, as the hydrogen atom returns to the dioxane linker upon structure relaxation rather than forming a radical state on the linker. Considering both AIMD trajectories and reaction barrier calculations, [Na(taci)][e$^-$] and [Na(1,4-dioxane)$_3$][e$^-$] would likely be resistant to deprotonation events at room temperature.

*Thermodynamic Stability of MOF Electride Candidates*

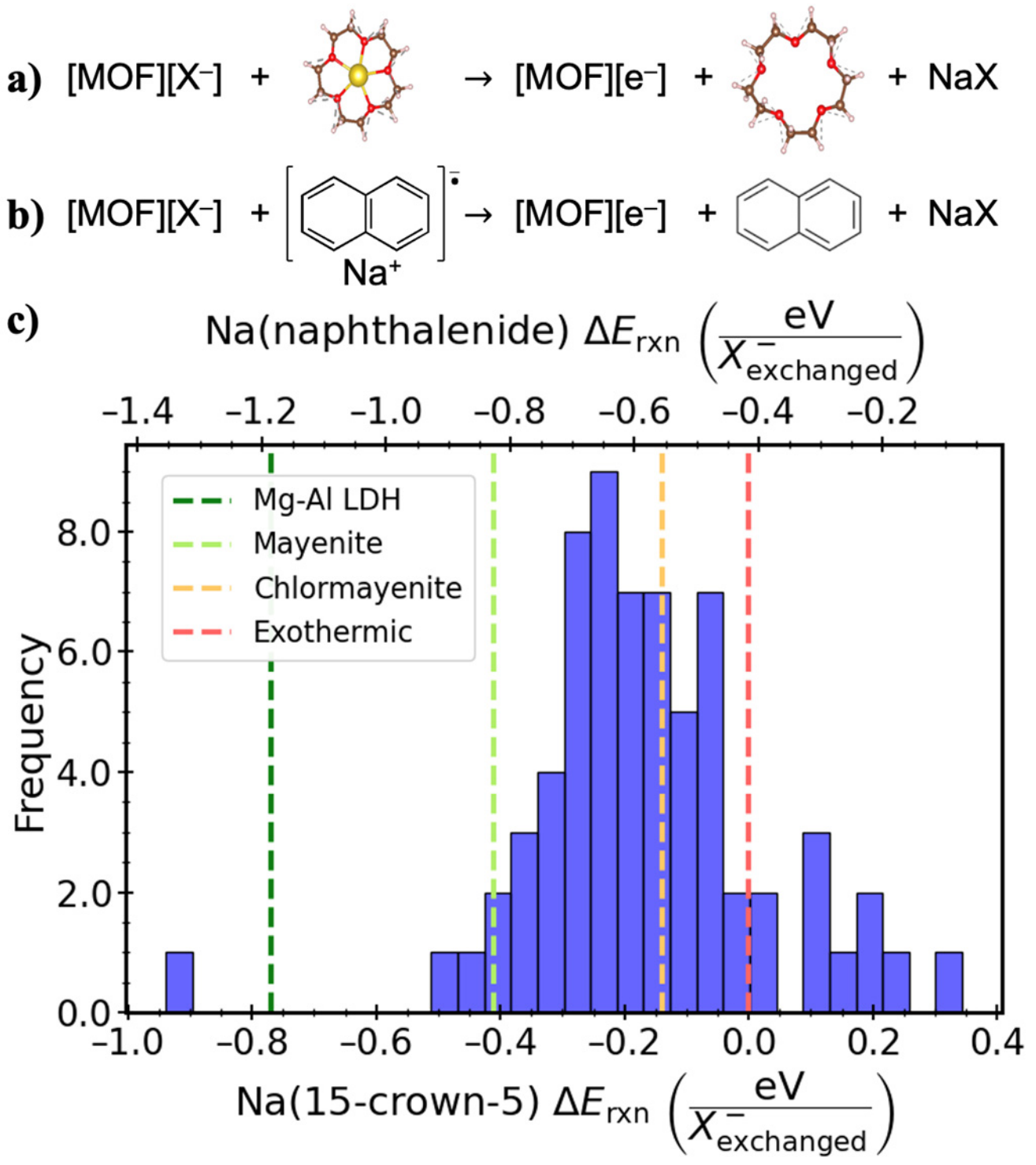

**Figure 6.** (a, b) The reaction of a cationic MOF ($X^-$ = halide) with a Na(15-crown-5) complex or Na(naphthalenide) to form the activated MOF electride. Atom color key: Na (gold), C (brown), H (pink), O (red). (c) The distribution of $\Delta E_{rxn}$ values corresponding to the activation of 67 MOF electride candidates via anion–electron exchange. The bottom axis displays the reaction energy when using Na(15-crown-5) as the reducing agent, whereas the top axis displays the reaction energy when using Na(naphthalenide). Dashed colored lines highlight points of reference, including $\Delta E_{rxn}$ values for known electrides.

The MOF electride candidates considered in this study represent the activated versions of previously synthesized MOFs. Experimentally, the synthesis of activated electrides involves the exchange of charge-balancing anions with electrons through a variety of possible reduction reactions.[1] One reported method to synthesize an activated electride is via a room-temperature, solution-phase process that uses Na(15-crown-5) in tetrahydrofuran (THF),[46] as depicted in Figure 6a. In this reaction, a charge-balancing halide can be exchanged with a solvated electron to form a sodium salt and the desired electride product.[46] While the ideal synthesis route will invariably depend on the particular material of interest, we have predicted the reaction energy, $\Delta E_{rxn}$, associated with both Na(15-crown-5) and a stronger Na(naphthalenide) reagent as two prototypical examples (Figures 6a and 6b). The use of an alternate reducing agent changes the magnitude of each reaction energy but does not alter the relative ordering of $\Delta E_{rxn}$ values among different electride candidates.

When using Na(15-crown-5), the median $\Delta E_{rxn}$ value is –0.20 eV per anion-exchanged, which suggests that this approach is mildly exothermic for most of the MOF electride candidates (Figure 6c). For comparison, we also show the $\Delta E_{rxn}$ values for the use of Na(naphthalenide), which increases the thermodynamic driving force for anion–electron exchange but may be too harsh of a reducing agent to maintain structural integrity for some MOF candidates. We note that air-free conditions are likely necessary to preserve the electride state during synthesis. While all MOFs are metastable due to their intrinsic porosity,[61] an energy above hull analysis[62] for each of the 67 MOF electride candidates indicates that they have a degree of metastability comparable to known, synthesized MOFs in the literature (Figure S9).[63]

As additional points of reference, we included the $\Delta E_{rxn}$ values for three synthesized electrides derived from a Mg–Al layered double hydroxide (LDH) (with $Cl^-$ interlayer anions),[46,64] mayenite[65] (with $O^{2-}$ anions), and chlormayenite[12] (with $Cl^-$ anions). Unlike the Mg–Al LDH electride, which has been synthesized using Na(15-crown-5),[46] the activated electride form of (chlor)mayenite has not been synthesized with either sodium-based reducing agent highlighted in Figure 6. However, we chose to include these prototypical electrides as additional reference points for $\Delta E_{rxn}$ values that are plausible in magnitude. We also note that while we have demonstrated that forming the electride state is thermodynamically feasible, the kinetics for the anion–electron exchange process are currently unknown. As such, it may be the case that the anion–electron exchange is sluggish for some of the MOF electride candidates, as has been observed for LDH electrides.[46]

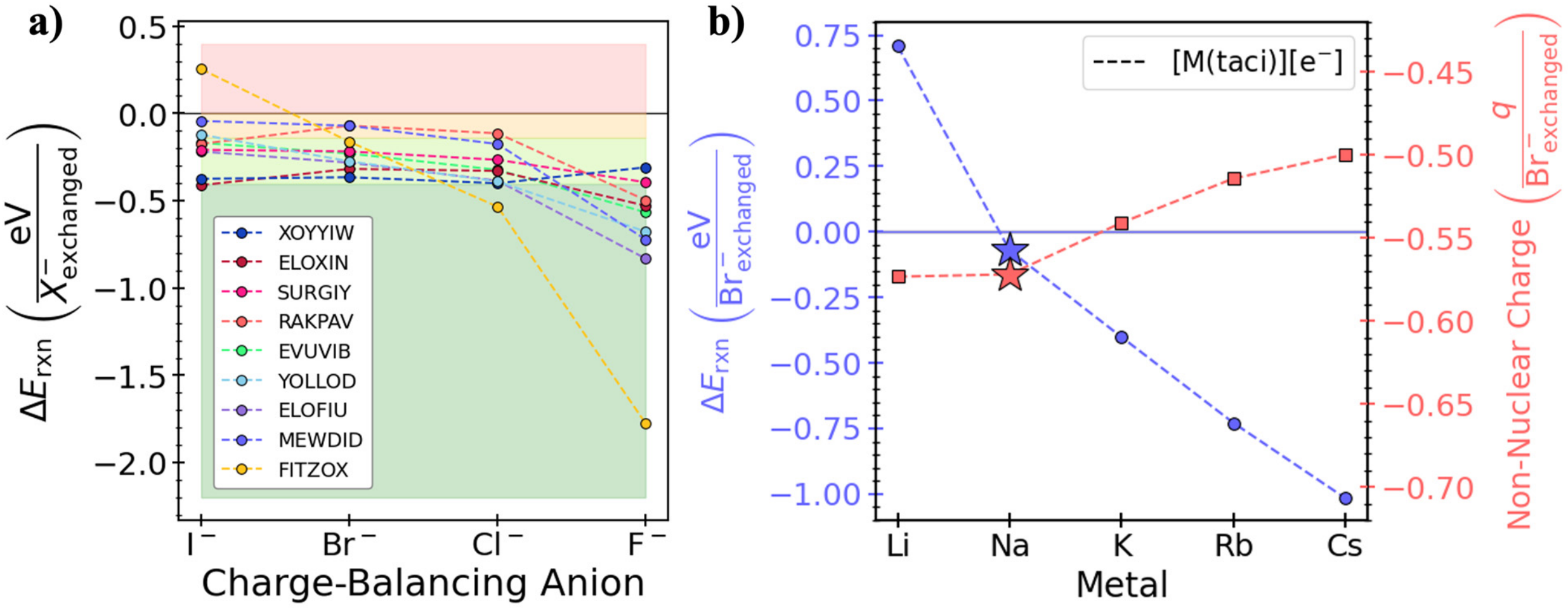

**Figure 7**. (a) Reaction energies ($\Delta E_{rxn}$) associated with the activation process (using the Na(15-crown-5) scheme) via anion–electron exchange for nine 3D-connected MOF electride candidates, including [Na(taci)][e⁻] (Refcode: MEWDID) and [Na(1,4-dioxane)$_3$][e⁻] (Refcode: FITZOX). These reaction energies were calculated with the originally reported anions as well as with other halides. Shaded regions are used to guide the eye based on the boundaries defined in Figure 6c. (b) Change in $\Delta E_{rxn}$ value and non-nuclear charge for [M(taci)][e⁻] (M = Li, Na, K, Rb, Cs), using $Br^-$ for the anion–electron exchange process. The as-synthesized MOF (i.e. Na) is denoted with a star.

To better understand which factors dictate the thermodynamic favorability of anion–electron exchange, we took the nine 3D-connected MOF electride candidates and varied the identity of the charge-balancing anions in the original crystal structure. For context, prior work has already demonstrated that the anions (and cations) in the $[Na(1,4\text{-dioxane})_3][X^-]$ family can be readily exchanged.[66,67] As shown in Figure 7a, the anion–electron exchange process is predicted to be more thermodynamically favorable when the charge-balancing anion is more electronegative. Therefore, on average, MOFs with $F^-$ anions are likely to be the most thermodynamically favorable candidates for anion–electron exchange among the halides. Given the relatively small size of the $F^-$ anion, these frameworks may also exhibit the lowest diffusion barriers for anion–electron exchange.

Given the synthetically tunable nature of MOFs, one can also consider exchanging the metal cations of the inorganic nodes to alter the thermodynamic favorability of electride formation. To identify structure–stability trends, we systematically varied the metal identity within $[Na(taci)][e^-]$ across the alkali metal family. As shown in Figure 7b, the more electropositive metals (e.g. $Cs^+$) result in a greater thermodynamic driving force towards anion–electron exchange. Across the alkali metals, there is a tradeoff between thermodynamic favorability of electride formation and the amount of non-nuclear charge (Figure 7b). However, we note that the variation in non-nuclear charge is relatively small, and the isostructural $[M(taci)][e^-]$ (M = Li, Na, K, Rb, Cs) frameworks are each predicted to have electride character, although the Li-containing variant is highly unlikely to be synthesizable based on the predicted thermochemistry.

*Bond Activation Enabled by the Pore-Confined Electrons*

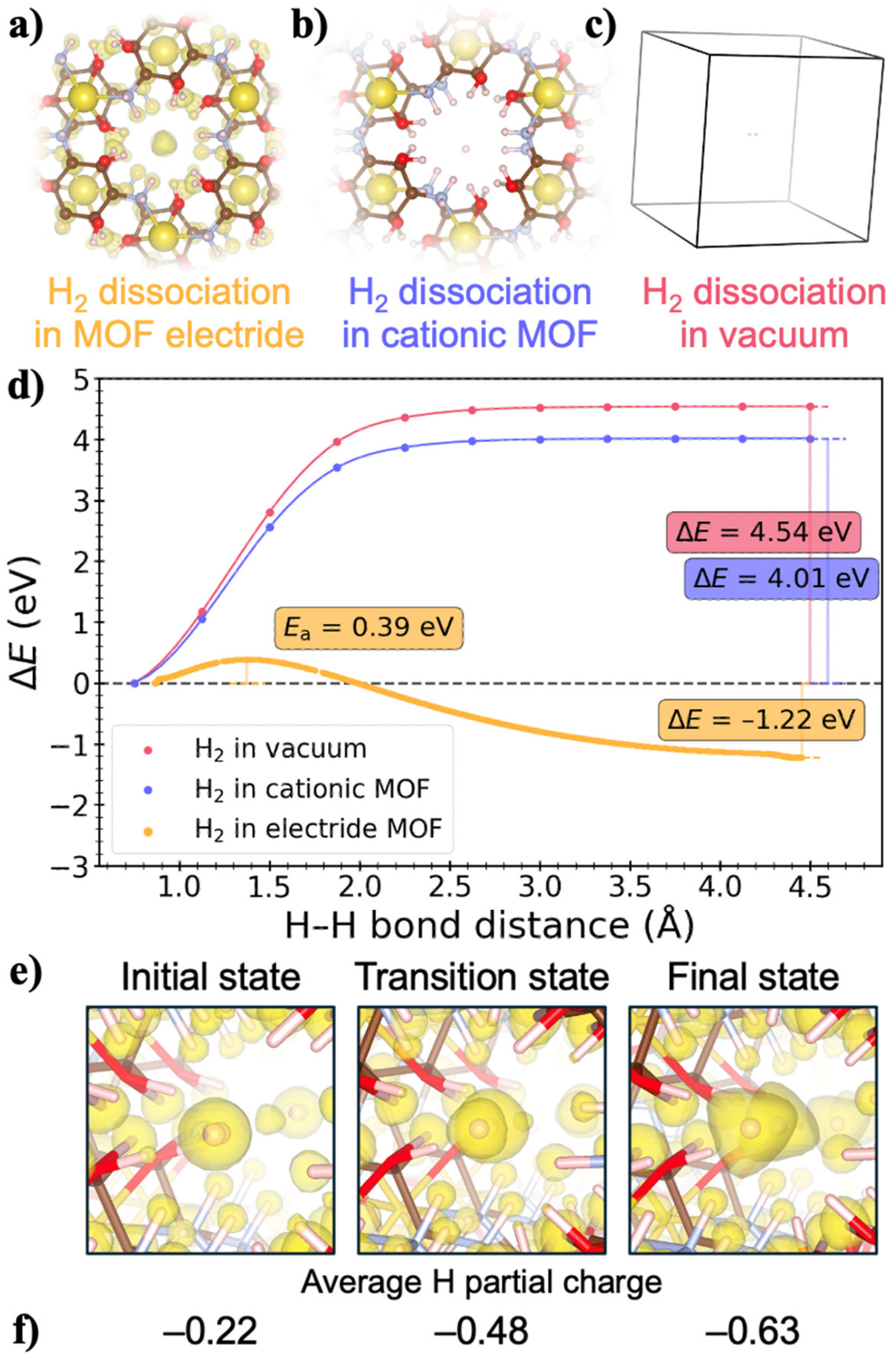

**Figure 8.** Depiction of an $H_2$ molecule in: (a) the center of the electride [Na(taci)][$e^-$] (ELF isosurface = 0.95), (b) the cationic MOF $[Na(taci)]^+$, and (c) vacuum. (d) Reaction pathway for $H_2$ dissociation in the aforementioned systems. (e) ELF visualization for the initial, transition, and final states of $H_2$ dissociation in the pore of the MOF electride (ELF isosurface = 0.98). (f) Average partial charge on each H atom during the reaction. Atom color key: Na (gold); C (brown); O (red); H (pink); N (light blue). All data in this figure represents calculations done at the PBE-D3(BJ) level of theory.

With the electron density that resides in the pore channels, a MOF electride would—in principle—be able to support a fundamentally new kind of catalytic mechanism in which charge transfer and bond-dissociation events can occur without adsorption at a surface site. As a proof-of-concept, we have chosen to study the dissociation of $H_2$ in the electron-rich pore space of [Na(taci)][$e^-$] to investigate the reactivity of the electride state. For comparison, we also computed the dissociation of $H_2$ in vacuum and in the pores of $[Na(taci)]^+$, wherein a charged calculation was carried out that eliminates the electride state.

Upon adsorption of $H_2$ in the pore of the electride MOF, the $H_2$ bond length increased from 0.75 Å (i.e. $H_2$ bond length in vacuum) to 0.86 Å, indicating that the H–H bond has weakened substantially in response to the electron-rich pore space. The effect of the electride state on the dissociation of $H_2$—as reflected in both the reaction energy, $\Delta E$, and activation energy, $E_a$—is quite pronounced (Figure 8d). The $\Delta E$ of reaction decreases by 5.24 eV and 5.76 eV with respect to $H_2$ dissociation in the cationic MOF and in vacuum—fundamentally changing from an endothermic to an exothermic process. Additionally, the intrinsic activation barrier for hydrogen dissociation in the electride MOF is reduced to a mere 0.39 eV. Single-point calculations at the HSE06-D3(BJ) level of theory were run on the initial, final, and transition states of $H_2$ dissociation in [Na(taci)][$e^-$], yielding $E_a = 0.38$ eV and $\Delta E = -1.45$ eV.

As the reaction progresses, electron density is donated to the hydrogen species from the electride state, as indicated by a visualization of the ELF (Figure 8e) and the increasingly negative partial charges on the H atoms (Figure 8f). As a result, the dissociated H atoms obtain significant hydride character, reducing the overall energy of the system. Since the $H_2$ dissociation process formally produces two $H^-$ as opposed to two H• species, both the activation energy and reaction energy are made more favorable than the non-electride version of the same MOF. We note that while the largest cavity diameter of [Na(taci)][$e^-$] is predicted to be 3.1 Å, the predicted pore-limiting diameter of 2.1 Å is smaller than the kinetic diameter of $H_2$ under the assumption of a rigid framework; nonetheless, this proof-of-concept is intended to demonstrate the new kinds of chemical reactivity that the pore-confined electrons of a MOF electride may offer.

It is important to emphasize that we have only focused on a single elementary step for the sake of demonstration. As realized here, $H_2$ dissociation in an electride MOF would be inherently non-catalytic since it is not a closed cycle (i.e. the pore-confined electrons would act as powerful stoichiometric reductants). That said, in analogy with the proposed reaction mechanism for ammonia synthesis on mayenite electride-supported ruthenium,[13] whereby the material reversibly stores hydrogen as $H^-$ and regenerates the electride state when the neutral product is formed, it may be possible for such an anionic electron to be regenerated when considered as part of a larger catalytic cycle or chemical looping process. It is also plausible that the pore-confined electrons of MOF electrides can facilitate the activation of other molecules with strong chemical bonds, such as $N_2$ or $CO_2$.

## Conclusion

In this work, we use *ab initio* calculations to establish and characterize MOF electrides: a new class of materials in which excess electrons are stabilized within the framework pore space. Through a high-throughput computational screening process, we identify multiple MOF candidates that are predicted to exhibit electride character following an anion–electron exchange process, analogous to the synthesis of known activated electrides in the literature. Leveraging the tunable nature of reticular chemistry, we show how systematic variation of the node composition and charge-balancing anions can be used to tailor the favorability of electride formation. Our calculations indicate that the use of electropositive, redox-inactive metal centers and highly electronegative charge-balancing anions favors electride formation.

By combining permanent porosity with spatially accessible, pore-confined electrons, MOF electrides introduce a fundamentally new mode of chemical reactivity that is not available in conventional materials. Our proof-of-concept calculations show that the pore-confined electrons can directly transfer charge to guest molecules and can cleave strong chemical bonds, as demonstrated for the dissociation of $H_2$. If experimentally realized, MOF electrides would offer a platform for a new kind of heterogeneous catalyst in which reactions can take place in the pore space, driven directly by electrons rather than atom-based adsorption sites. The discovery of permanently porous materials with pore-confined electrons may offer a unique opportunity to circumvent known scaling relationships in catalysis[68] by decoupling chemical reactivity from chemisorption, while transforming otherwise empty void space into a new kind of active site altogether.

## Resource Availability

The CSD Refcodes and tabulated data associated with the screening process are hosted on Figshare at the following DOI: 10.6084/m9.figshare.33648082. The DFT calculations from this work will be hosted on NOMAD[69] upon article acceptance at the following DOI: <placeholder>.

## Acknowledgments

The authors gratefully acknowledge financial support from the Breakthrough Energy Foundation and the Princeton Laboratory for Artificial Intelligence's $AI^2$ Initiative. The simulations presented in this article were performed on computational resources managed and supported by Princeton University's Research Computing.

## Author Contributions

J.H.B.: Data curation, Formal analysis, Investigation, Methodology, Validation, Visualization, Writing – original draft, Writing – review and editing. A.S.R.: Conceptualization, Funding acquisition, Investigation, Methodology, Project administration, Resources, Supervision, Writing – review and editing.

## Declaration of Interests

The authors declare no competing interests.

## Supplemental Information

Document S1. Supplemental methods, Figures S1–S12, and Table S1.

## References

(1) Liu, C.; Nikolaev, S. A.; Ren, W.; Burton, L. Electrides: A Review. *J. Mater. Chem. C* **2020**, *8* (31), 10551–10567. https://doi.org/10.1039/D0TC01165G

(2) Hosono, H.; Kitano, M. Advances in Materials and Applications of Inorganic Electrides. *Chem Rev* **2021**, *121* (5), 3121–3185.

(3) Druffel, D. L.; Woomer, A. H.; Kuntz, K. L.; Pawlik, J. T.; Warren, S. C. Electrons on the Surface of 2D Materials: From Layered Electrides to 2D Electrenes. *J. Mater. Chem. C* **2017**, *5* (43), 11196–11213. https://doi.org/10.1039/C7TC02488F

(4) Inoshita, T.; Saito, S.; Hosono, H. Floating Interlayer and Surface Electrons in 2D Materials: Graphite, Electrides, and Electrenes. *Small Sci.* **2021**, *1* (9), 2100020. https://doi.org/10.1002/smsc.202100020

(5) Dye, J. L. Electrides: Ionic Salts with Electrons as the Anions. *Science* **1990**, *247* (4943), 663–668. https://doi.org/10.1126/science.247.4943.663

(6) Dye, J. L. Electrides: Early Examples of Quantum Confinement. *Acc. Chem. Res.* **2009**, *42* (10), 1564–1572. https://doi.org/10.1021/ar9000857

(7) Davison, N.; Quirk, J. A.; Tuna, F.; Collison, D.; McMullin, C. L.; Michaels, H.; Morritt, G. H.; Waddell, P. G.; Gould, J. A.; Freitag, M.; Dawson, J. A.; Lu, E. A Room-Temperature-Stable Electride and Its Reactivity: Reductive Benzene/Pyridine Couplings and Solvent-Free Birch Reductions. *Chem* **2023**, *9* (3), 576–591. https://doi.org/10.1016/j.chempr.2022.11.006

(8) Redko, M. Y.; Jackson, J. E.; Huang, R. H.; Dye, J. L. Design and Synthesis of a Thermally Stable Organic Electride. *J. Am. Chem. Soc.* **2005**, *127* (35), 12416–12422. https://doi.org/10.1021/ja053216f

(9) Day, C. S.; Do, C. D.; Odena, C.; Benet-Buchholz, J.; Xu, L.; Foroutan-Nejad, C.; Hopmann, K. H.; Martin, R. Room-Temperature-Stable Magnesium Electride via Ni(II) Reduction. *J. Am. Chem. Soc.* **2022**, *144* (29), 13109–13117. https://doi.org/10.1021/jacs.2c01807

(10) Bowles, A. W. J.; Quirk, J. A.; Liu, Y.; Morritt, G. H.; Freitag, M.; Whitehead, G. F. S.; Woodward, A. W.; Brookfield, A.; Goodwin, C. A. P.; Collison, D.; Tuna, F.; McMullin, C. L.; Dawson, J. A.; Lu, E.; Ortu, F. Mechanochemical Synthesis, Characterization and Reactivity of a Room Temperature Stable Calcium Electride. *J. Am. Chem. Soc.* **2024**, *146* (42), 28914–28924. https://doi.org/10.1021/jacs.4c09408

(11) Matsuishi, S.; Toda, Y.; Miyakawa, M.; Hayashi, K.; Kamiya, T.; Hirano, M.; Tanaka, I.; Hosono, H. High-Density Electron Anions in a Nanoporous Single Crystal: $[Ca_{24}Al_{28}O_{64}]^{4+}(4e^-)$. *Science* **2003**, *301* (5633), 626–629. https://doi.org/10.1126/science.1083842

(12) Wang, R.; Yang, H.; Lu, Y.; Kanamori, K.; Nakanishi, K.; Guo, X. Synthesis, Reduction, and Electrical Properties of Macroporous Monolithic Mayenite Electrides with High Porosity. *ACS Omega* **2017**, *2* (11), 8148–8155. https://doi.org/10.1021/acsomega.7b01121

(13) Kitano, M.; Inoue, Y.; Yamazaki, Y.; Hayashi, F.; Kanbara, S.; Matsuishi, S.; Yokoyama, T.; Kim, S.-W.; Hara, M.; Hosono, H. Ammonia Synthesis Using a Stable Electride as an Electron Donor and Reversible Hydrogen Store. *Nat. Chem.* **2012**, *4* (11), 934–940. https://doi.org/10.1038/nchem.1476

(14) Inoue, Y.; Kitano, M.; Kim, S.-W.; Yokoyama, T.; Hara, M.; Hosono, H. Highly Dispersed Ru on Electride $[Ca_{24}Al_{28}O_{64}]^{4+}(e^{-})_4$ as a Catalyst for Ammonia Synthesis. *ACS Catal.* **2014**, *4* (2), 674–680. https://doi.org/10.1021/cs401044a

(15) Hara, M.; Kitano, M.; Hosono, H. Ru-Loaded C12A7:e– Electride as a Catalyst for Ammonia Synthesis. *ACS Catal.* **2017**, *7* (4), 2313–2324. https://doi.org/10.1021/acscatal.6b03357

(16) Yaghi, O. M.; Kalmutzki, M. J.; Diercks, C. S. *Introduction to Reticular Chemistry*; Wiley, 2019. https://doi.org/10.1002/9783527821099

(17) Calbo, J.; Golomb, M. J.; Walsh, A. Redox-Active Metal–Organic Frameworks for Energy Conversion and Storage. *J. Mater. Chem. A* **2019**, *7* (28), 16571–16597. https://doi.org/10.1039/C9TA04680A

(18) Xie, L. S.; Skorupskii, G.; Dincă, M. Electrically Conductive Metal–Organic Frameworks. *Chem. Rev.* **2020**, *120* (16), 8536–8580. https://doi.org/10.1021/acs.chemrev.9b00766

(19) Ryu, J.; Sippach, L.; Hallweger, S. A.; Palatinus, L.; Müller, P.; Glazyrin, K.; Kieslich, G.; Oppenheim, J. J.; Dincă, M. Evidence for an Electronically Driven Charge Density Wave in a 1D Metallic MOF. *ACS Cent. Sci.* **2026**, *12* (5), 704–711. https://doi.org/10.1021/acscentsci.6c00405

(20) Goesten, M. G.; Schoop, L. M. Diradicals as Topological Charge Carriers in Metal–Organic Toy Model Pt3(HIB)2. *J. Am. Chem. Soc.* **2024**, *146* (43), 29599–29608. https://doi.org/10.1021/jacs.4c09993

(21) Csányi, G.; Littlewood, P. B.; Nevidomskyy, A. H.; Pickard, C. J.; Simons, B. D. The Role of the Interlayer State in the Electronic Structure of Superconducting Graphite Intercalated Compounds. *Nat. Phys.* **2005**, *1* (1), 42–45. https://doi.org/10.1038/nphys119

(22) Berg, L.; Goesten, M. G. Quasiatomic Pore Bands. *Newton* **2026**, 100635. https://doi.org/10.1016/j.newton.2026.100635

(23) Golomb, M. J.; Tolborg, K.; Calbo, J.; Walsh, A. Role of Counterions in the Structural Stabilisation of Redox-Active Metal-Organic Frameworks. *Chem. – Eur. J.* **2023**, *29* (16), e202203843. https://doi.org/10.1002/chem.202203843

(24) Burton, L. A. Best Practices for Modelling Electrides. arXiv 2025. https://doi.org/10.48550/ARXIV.2512.24989

(25) Weaver, S. M.; Lanetti, M. G.; Slamowitz, C. C.; Radomsky, R. C.; Warren, S. C. Assessing Dimensionality in Electrides. *J. Phys. Chem. C* **2025**, *129* (6), 3211–3217. https://doi.org/10.1021/acs.jpcc.4c06803

(26) Dale, S. G.; Otero-de-la-Roza, A.; Johnson, E. R. Density-Functional Description of Electrides. *Phys Chem Chem Phys* **2014**, *16* (28), 14584–14593. https://doi.org/10.1039/C3CP55533J

(27) Dale, S. G.; Johnson, E. R. Theoretical Descriptors of Electrides. *J. Phys. Chem. A* **2018**, *122* (49), 9371–9391. https://doi.org/10.1021/acs.jpca.8b08548

(28) Weaver, S. M.; Sundberg, J. D.; Slamowitz, C. C.; Radomsky, R. C.; Lanetti, M. G.; McRae, L. M.; Warren, S. C. Counting Electrons in Electrides. *J. Am. Chem. Soc.* **2023**, *145* (48), 26472–26476. https://doi.org/10.1021/jacs.3c10876

(29) Meng, W.; Bai, J.; Ma, F.; Jiao, Y.; Wang, S.; Jiang, J.; Zhang, X.; Cheng, Z.; Yang, T. Magnetic Electrides: Recent Advances in Materials Realization and Application Prospects. *Appl. Phys. Rev.* **2025**, *12* (1), 011320. https://doi.org/10.1063/5.0233622

(30) Tada, T.; Takemoto, S.; Matsuishi, S.; Hosono, H. High-Throughput Ab Initio Screening for Two-Dimensional Electride Materials. *Inorg. Chem.* **2014**, *53* (19), 10347–10358. https://doi.org/10.1021/ic501362b

(31) Inoshita, T.; Jeong, S.; Hamada, N.; Hosono, H. Exploration for Two-Dimensional Electrides via Database Screening and *Ab Initio* Calculation. *Phys. Rev. X* **2014**, *4* (3), 031023. https://doi.org/10.1103/PhysRevX.4.031023
(32) Zhang, Y.; Wang, H.; Wang, Y.; Zhang, L.; Ma, Y. Computer-Assisted Inverse Design of Inorganic Electrides. *Phys. Rev. X* **2017**, *7* (1), 011017. https://doi.org/10.1103/PhysRevX.7.011017
(33) Burton, L. A.; Ricci, F.; Chen, W.; Rignanese, G.-M.; Hautier, G. High-Throughput Identification of Electrides from All Known Inorganic Materials. *Chem. Mater.* **2018**, *30* (21), 7521–7526. https://doi.org/10.1021/acs.chemmater.8b02526
(34) Zhu, Q.; Frolov, T.; Choudhary, K. Computational Discovery of Inorganic Electrides from an Automated Screening. *Matter* **2019**, *1* (5), 1293–1303. https://doi.org/10.1016/j.matt.2019.06.017
(35) Zhou, J.; Shen, L.; Yang, M.; Cheng, H.; Kong, W.; Feng, Y. P. Discovery of Hidden Classes of Layered Electrides by Extensive High-Throughput Material Screening. *Chem. Mater.* **2019**, *31* (6), 1860–1868. https://doi.org/10.1021/acs.chemmater.8b03021
(36) Wang, Z.; Gong, Y.; Evans, M. L.; Yan, Y.; Wang, S.; Miao, N.; Zheng, R.; Rignanese, G.-M.; Wang, J. Machine Learning-Accelerated Discovery of $A_2BC_2$ Ternary Electrides with Diverse Anionic Electron Densities. *J. Am. Chem. Soc.* **2023**, *145* (48), 26412–26424. https://doi.org/10.1021/jacs.3c10538
(37) Zhang, X.; Meng, W.; Liu, Y.; Dai, X.; Liu, G.; Kou, L. Magnetic Electrides: High-Throughput Material Screening, Intriguing Properties, and Applications. *J. Am. Chem. Soc.* **2023**, *145* (9), 5523–5535. https://doi.org/10.1021/jacs.3c00284
(38) Hirayama, M.; Matsuishi, S.; Hosono, H.; Murakami, S. Electrides as a New Platform of Topological Materials. *Phys. Rev. X* **2018**, *8* (3), 031067. https://doi.org/10.1103/PhysRevX.8.031067
(39) McRae, L. M.; Radomsky, R. C.; Pawlik, J. T.; Druffel, D. L.; Sundberg, J. D.; Lanetti, M. G.; Donley, C. L.; White, K. L.; Warren, S. C. $Sc_2$ C, a 2D Semiconducting Electride. *J. Am. Chem. Soc.* **2022**, *144* (24), 10862–10869. https://doi.org/10.1021/jacs.2c03024
(40) Horiba, K.; Yukawa, R.; Mitsuhashi, T.; Kitamura, M.; Inoshita, T.; Hamada, N.; Otani, S.; Ohashi, N.; Maki, S.; Yamaura, J.; Hosono, H.; Murakami, Y.; Kumigashira, H. Semimetallic Bands Derived from Interlayer Electrons in the Quasi-Two-Dimensional Electride Y 2 C. *Phys. Rev. B* **2017**, *96* (4), 045101. https://doi.org/10.1103/PhysRevB.96.045101
(41) Chanhom, P.; Fritz, K. E.; Burton, L. A.; Kloppenburg, J.; Filinchuk, Y.; Senyshyn, A.; Wang, M.; Feng, Z.; Insin, N.; Suntivich, J.; Hautier, G. Sr3CrN3: A New Electride with a Partially Filled d-Shell Transition Metal. *J. Am. Chem. Soc.* **2019**, *141* (27), 10595–10598. https://doi.org/10.1021/jacs.9b03472
(42) Dong, X.; Oganov, A. R.; Goncharov, A. F.; Stavrou, E.; Lobanov, S.; Saleh, G.; Qian, G.-R.; Zhu, Q.; Gatti, C.; Deringer, V. L.; Dronskowski, R.; Zhou, X.-F.; Prakapenka, V. B.; Konôpková, Z.; Popov, I. A.; Boldyrev, A. I.; Wang, H.-T. A Stable Compound of Helium and Sodium at High Pressure. *Nat. Chem.* **2017**, *9* (5), 440–445. https://doi.org/10.1038/nchem.2716
(43) Yu, J.; Pei, C.; Hosono, H.; Blatov, V. A.; Qi, Y.; Wang, J. Discovery of Electron–Neutral Electrides with Pressure-Induced Superconductivity. *J. Am. Chem. Soc.* **2026**. https://doi.org/10.1021/jacs.6c13689
(44) Xiao, C.; Bristowe, N.; Mostofi, A. A. Theory and Discovery of Electrides. arXiv 2026. https://doi.org/10.48550/ARXIV.2605.11724
(45) Moghadam, P. Z.; Li, A.; Wiggin, S. B.; Tao, A.; Maloney, A. G. P.; Wood, P. A.; Ward, S. C.; Fairen-Jimenez, D. Development of a Cambridge Structural Database Subset: A Collection of Metal–Organic Frameworks for Past, Present, and Future. *Chem. Mater.* **2017**, *29* (7), 2618–2625. https://doi.org/10.1021/acs.chemmater.7b00441
(46) Kitano, M.; Hosono, H.; Yokoyama, T.; Ogasawara, K.; Agawa, Y. Layered Double Hydroxide Electride and Method for Producing Same. US12371336B2, July 29, 2025.
(47) Perdew, J. P.; Burke, K.; Ernzerhof, M. Generalized Gradient Approximation Made Simple. *Phys. Rev. Lett.* **1996**, *77* (18), 3865–3868. https://doi.org/10.1103/PhysRevLett.77.3865

(48) Becke, A. D.; Johnson, E. R. Exchange-Hole Dipole Moment and the Dispersion Interaction. *J. Chem. Phys.* **2005**, *122* (15), 154104. https://doi.org/10.1063/1.1884601
(49) Grimme, S.; Ehrlich, S.; Goerigk, L. Effect of the Damping Function in Dispersion Corrected Density Functional Theory. *J. Comput. Chem.* **2011**, *32* (7), 1456–1465. https://doi.org/10.1002/jcc.21759
(50) Grimme, S.; Antony, J.; Ehrlich, S.; Krieg, H. A Consistent and Accurate Ab Initio Parametrization of Density Functional Dispersion Correction (DFT-D) for the 94 Elements H-Pu. *J. Chem. Phys.* **2010**, *132* (15), 154104. https://doi.org/10.1063/1.3382344
(51) Novoselov, D. Y.; Mazannikova, M. A.; Korotin, D. M.; Shorikov, A. O.; Korotin, M. A.; Anisimov, V. I.; Oganov, A. R. Localization Mechanism of Interstitial Electronic States in Electride Mayenite. *J. Phys. Chem. Lett.* **2022**, *13* (31), 7155–7160. https://doi.org/10.1021/acs.jpclett.2c02002
(52) Kang, B.; Parrish, K.; Zhu, Q. First-Principles Investigation of Electrides Derived from Sodalites. *J. Phys. Chem. C* **2023**, *127* (37), 18745–18754. https://doi.org/10.1021/acs.jpcc.3c04681
(53) Savin, A.; Nesper, R.; Wengert, S.; Fässler, T. F. ELF: The Electron Localization Function. *Angew. Chem. Int. Ed. Engl.* **1997**, *36* (17), 1808–1832. https://doi.org/10.1002/anie.199718081
(54) Mori-Sánchez, P.; Cohen, A. J.; Yang, W. Many-Electron Self-Interaction Error in Approximate Density Functionals. *J. Chem. Phys.* **2006**, *125* (20), 201102. https://doi.org/10.1063/1.2403848
(55) Bryenton, K. R.; Adeleke, A. A.; Dale, S. G.; Johnson, E. R. Delocalization Error: The Greatest Outstanding Challenge in Density-functional Theory. *WIREs Comput. Mol. Sci.* **2023**, *13* (2), e1631. https://doi.org/10.1002/wcms.1631
(56) Heyd, J.; Scuseria, G. E.; Ernzerhof, M. Hybrid Functionals Based on a Screened Coulomb Potential. *J. Chem. Phys.* **2003**, *118* (18), 8207–8215. https://doi.org/10.1063/1.1564060
(57) Krukau, A. V.; Vydrov, O. A.; Izmaylov, A. F.; Scuseria, G. E. Influence of the Exchange Screening Parameter on the Performance of Screened Hybrid Functionals. *J. Chem. Phys.* **2006**, *125* (22), 224106. https://doi.org/10.1063/1.2404663
(58) Reiss, G. J.; Hegetschweiler, K. Poly[[($\mu_4$ -1,3,5-Triamino-1,3,5-Trideoxy- *Cis* -Inositol)Sodium] Bromide]. *Acta Crystallogr. Sect. E Struct. Rep. Online* **2013**, *69* (4), m185–m186. https://doi.org/10.1107/S1600536813005618
(59) Belsky, V. K.; Maslennikova, V. I. Sodium Tris(1,4-Dioxane) Iodide. *Acta Crystallogr. C* **1999**, *55* (6), IUC9900057. https://doi.org/10.1107/S0108270199099370
(60) Xie, Y.; Zhang, S.-Y.; Yin, Y.; Zheng, N.; Ali, A.; Younis, M.; Ruan, S.; Zeng, Y.-J. Emerging Ferromagnetic Materials for Electrical Spin Injection: Towards Semiconductor Spintronics. *Npj Spintron.* **2025**, *3* (1), 10. https://doi.org/10.1038/s44306-024-00070-z
(61) Akimbekov, Z.; Navrotsky, A. Little Thermodynamic Penalty for the Synthesis of Ultraporous Metal Organic Frameworks. *ChemPhysChem* **2016**, *17* (4), 468–470. https://doi.org/10.1002/cphc.201501086
(62) Bartel, C. J. Review of Computational Approaches to Predict the Thermodynamic Stability of Inorganic Solids. *J. Mater. Sci.* **2022**, *57* (23), 10475–10498. https://doi.org/10.1007/s10853-022-06915-4
(63) Dallmann, B.; Saha, A.; Rosen, A. S. Predicting the Thermodynamic Limits of Metal–Organic Framework Metastability. *J. Am. Chem. Soc.* **2026**, *148* (19), 19487–19501. https://doi.org/10.1021/jacs.5c20253
(64) Shi, Y.; Rosen, A. S. Layered Double Hydroxides with Interlayer Electrons Activate Strong Chemical Bonds. *ChemRxiv* **2026**. https://doi.org/10.26434/chemrxiv.15008487/v1
(65) Khan, K.; Tareen, A. K.; Aslam, M.; Thebo, K. H.; Khan, U.; Wang, R.; Shams, S. S.; Han, Z.; Ouyang, Z. A Comprehensive Review on Synthesis of Pristine and Doped Inorganic Room Temperature Stable Mayenite Electride, [Ca24Al28O64]4+(e−)4 and Its Applications as a Catalyst. *Prog. Solid State Chem.* **2019**, *54*, 1–19. https://doi.org/10.1016/j.progsolidstchem.2018.12.001
(66) Chen, X.; Qin, Y.; Weng, Y.; Song, X.; Lv, H.; Liao, W.; Xiong, R. Neutral X-Site $ABX_3$ -Type Perovskites. *Angew. Chem. Int. Ed.* **2026**, *65* (28), e3168284. https://doi.org/10.1002/anie.3168284

(67) Chen, X.-G.; Qin, Y.; Lv, H.-P.; Song, X.-J.; Liao, W.-Q.; Xiong, R.-G. Metal–Organic Framework Goes Perovskite: A Self-Healing Neutral X-Site Perovskite Ferroelastic Crystal. *J. Am. Chem. Soc.* **2026**, *148* (22), 23383–23391. https://doi.org/10.1021/jacs.6c08330
(68) Pérez-Ramírez, J.; López, N. Strategies to Break Linear Scaling Relationships. *Nat. Catal.* **2019**, *2* (11), 971–976. https://doi.org/10.1038/s41929-019-0376-6
(69) Scheidgen, M.; Himanen, L.; Ladines, A. N.; Sikter, D.; Nakhaee, M.; Fekete, Á.; Chang, T.; Golparvar, A.; Márquez, J. A.; Brockhauser, S.; Brückner, S.; Ghiringhelli, L. M.; Dietrich, F.; Lehmberg, D.; Denell, T.; Albino, A.; Näsström, H.; Shabih, S.; Dobener, F.; Kühbach, M.; Mozumder, R.; Rudzinski, J. F.; Daelman, N.; Pizarro, J. M.; Kuban, M.; Salazar, C.; Ondračka, P.; Bungartz, H.-J.; Draxl, C. NOMAD: A Distributed Web-Based Platform for Managing Materials Science Research Data. *J. Open Source Softw.* **2023**, *8* (90), 5388. https://doi.org/10.21105/joss.05388

# Supplementary Information

## Towards a Metal–Organic Framework with Pore-Confined Electrons

Julia H. Baratta,[a] Andrew S. Rosen[a,*]

[a]Department of Chemical and Biological Engineering, Princeton University, Princeton, NJ 08544, United States

[*]Corresponding author: asrosen@princeton.edu

## Methods

### *Density Functional Theory Calculations*

Density functional theory (DFT) calculations were carried out using the Vienna *Ab initio* Simulation Package (VASP) v6.5.1.[1,2] The VASP calculations were orchestrated using Quantum Accelerator (QuAcc)[3] v1.3.2, which is built upon the Atomic Simulation Environment.[4] Jobflow[5] and Jobflow-Remote were used to dispatch and monitor the calculations. The DFT calculations in the high-throughput screening process were carried out using the Perdew–Burke–Ernzerhof (PBE)[6] functional with D3(BJ)[7–9] dispersion corrections. Relaxations of the unit cell and atomic positions were run with a plane-wave kinetic energy cutoff of 520 eV, projector-augmented-wave (PAW)[10] v.64 VASP-recommended PBE pseudopotentials, a *k*-spacing of 0.4 $Å^{-1}$, a maximum force tolerance of 0.02 eV/Å, and a self-consistent field (SCF) convergence criterion of $10^{-6}$ eV. Non-spherical contributions to the gradient corrections in the PAW spheres were included, and the precision was set to "accurate." Static calculations, used to generate electronic structure data, were run on these relaxed structures with the same calculation settings unless otherwise stated. All calculations were considered with spin polarization and the default, high-spin magnetic initialization in QuAcc. SeeK-path v2.2.0[11,12] was used to generate a *k*-path (enumeration of high-symmetry points in a reciprocal lattice) and associated primitive cell for the band structure calculations. Additional parameters can be found in the `RosenSetPBE` VASP preset in QuAcc. The crystal structures and volumetric data were visualized using VESTA.[13]

Brief *ab initio* molecular dynamics (AIMD) simulations were carried out in the *NVT* ensemble at 300 K as implemented in QuAcc v1.5.5 at the PBE-D3(BJ) level of theory using the same DFT settings as listed above. In addition, a timestep of 0.5 fs was used, and the temperature was maintained using a Nosé–Hoover thermostat.[14] AIMD runs were given a maximum simulation duration of 10 ps, although the average simulation length was 1.26 ps since a calculation wall-time of 24 hours was adopted (distribution shown in Figure S4).

While PBE is widely adopted for studying electrides,[15] it is expected to systematically underpredict the degree of electron localization, which the inclusion of Hartree–Fock exchange can help resolve. As such, select static calculations were run with the screened hybrid functional HSE06[8,16,17] on geometries previously relaxed with PBE-D3(BJ). The remaining parameters generally correspond to those from the PBE-D3(BJ) calculations, except that a tighter *k*-spacing of 0.2 $Å^{-1}$ was adopted. Additional parameters can be found in the `DefaultSetHybrid` VASP preset in QuAcc.

### *CSD Queries*

The 100,559 Cambridge Structural Database (CSD) reference codes contained in version 6.01 of the non-disordered MOF subset of the CSD[18] were queried to find (1) cationic MOFs with charge-balancing anions and (2) MOFs with redox-inactive metals or metals in their lowest oxidation states. Specifically, the following queries were used to obtain 717 MOF reference codes which were then preprocessed according to the details outlined in the next section.

Using the "Build Queries" and "Combine Queries" sections of ConQuest,[19] we created two separate queries.

Halogen Anions

"must have":

1. Group 7A; charge = -1; Number of bonded atoms= 0

"must have at least one of":

1. 1A; charge = 0/+
2. 2A; charge = 0/2+
3. QA = Sc, Y; charge = 0/3+
4. QA = LN, AN; charge = 0/3+
5. Ag; charge = 0/+
6. Au; charge = 0/3+
7. 2B; charge = 0/+/2+
8. 3A; charge = 0/+/3+
9. 4A; charge = 0/2+/4+
10. As, Sb, Bi; charge = 0/3+
11. Te, Po; charge = 0/2+
12. Cu; charge = 0/+

Hydroxyl Anions

"must have":

1. O-H; charge = -1; Number of bonded atoms= 0

"must have at least one of":

1. 1A; charge = 0/+
2. 2A; charge = 0/2+
3. QA = Sc, Y, charge = 0/3+
4. QA = LN, AN; charge = 0/3+
5. Ag; charge = 0/+
6. Au; charge = 0/3+
7. 2B; charge = 0/+/2+
8. 3A; charge = 0/+/3+
9. 4A; charge = 0/2+/4+
10. As, Sb, Bi; charge = 0/3+
11. Te, Po; charge = 0/2+
12. Cu; charge = 0/+

*Preprocessing*

1. We used MOFChecker[20] v0.9.6 to perform the following structural checks:

```
"has_carbon": true,
"has_hydrogen": true,
"has_atomic_overlaps": false,
"has_overcoordinated_c": false,
"has_overcoordinated_n": false,
"has_overcoordinated_h": false,
"has_undercoordinated_n": false,
"has_undercoordinated_rare_earth": false,
"has_metal": true,
"has_suspicious_terminal_oxo": false,
"has_undercoordinated_alkali_alkaline": false,
```

`"has_geometrically_exposed_metal": false`

2. We used SAMOSA[21] for free solvent removal and anion removal
3. We created primitive cells with Pymatgen[22]
4. We eliminated structures with over 750 atoms per primitive cell
5. We established a porosity lower bound of largest cavity diameter of 2.4 Å
6. We excluded all MOF structures containing the elements Ce–Lu. VASP-recommended pseudopotentials of lanthanides Pr–Lu have a fixed oxidation state, which could influence the ground-state electronic structure.

The above steps left 314 DFT-ready primitive cells.

*Assessing lattice constants of activated structures*

To eliminate MOFs that had collapsed or substantially changed upon optimization after anion–electron exchange, we filtered out structures with a deviation of $\geq$15% between the computed lattice parameters and the values reported for the as-synthesized MOF in the Cambridge Structural Database.

## *Electronic Structure Analysis*

Critic2[23] v. 1.2.447 was used to identify non-nuclear, zero-flux basins in the ELF scalar field with a nearest-atom cutoff distance of 3.0 bohr, as justified based on Figures S2 and S3. Two features were evaluated: the ELF non-nuclear maxima and total non-nuclear charge. The non-nuclear charge was calculated by integrating charge density using the Yu–Trinkle[24] method over the zero-flux ELF basins and normalizing this total charge by the number of anions originally exchanged in the structure. Since all DFT calculations were spin polarized, both the ELF scalar field (ELFCAR) and charge density (CHGCAR) were split into their respective spin channels, integrated separately, and the density from each integration was summed to get the total non-nuclear charge density across both spin channels. Non-nuclear maxima were identified by recording the coordinates of the zero-flux basins and finding the maximum ELF value contained within these coordinates across both spin channels. ELF values are scaled between zero and one and, therefore, do not need to be normalized on a per-anion-exchanged basis.

## *Distinct Chemistry Assessment*

Duplicates within CSD Refcode families (i.e. same six-letter codes) were removed because they represented MOFs with functionally identical chemistries and connectivity. MOFs originally from the same paper (i.e. same first three letters) were manually inspected and eliminated if they were determined to be functionally identical.

## *Connectivity Assessment*

MOF electride candidates were split by dimensionality based on the designations made by the Cambridge Structural Database upon inclusion in the "non-disordered MOF subset" database.

Porosity Assessment

The pore limiting diameter and largest cavity diameters were calculated using Zeo++ (zeopp-lsmo) v0.4.7[25] for the 250 MOFs with converged relaxations after anion–electron exchange.

## *Thermodynamic Stability Calculations*

Reaction energies were calculated using the electronic energies from DFT relaxations based on the reaction scheme in Figures 6a and 6b. Sodiated and non-sodiated 15-crown-5 were relaxed as lone molecules in a 20 Å×20 Å×20 Å cell with an analogous procedure for the naphthalene. All sodium salts were relaxed from the lowest-energy structures on the Materials Project.[26,27] The unit cell and atomic positions of the original crystal structures (i.e. with anions) were relaxed from the experimentally reported structures on the CSD, following free solvent removal with SAMOSA (commit 32056ce).[21] Stability with respect to deprotonation was determined by identifying gas ($H_2$, $H_2O$, $CO$, $CO_2$, $CH_4$, $NH_3$, $N_2$, HCN) formation during the AIMD trajectory. Gas formation was detected using the JmolNN nearest-neighbor search method implemented in Pymatgen v. 2025.10.7[22] (tolerance parameter = 0.45 Å).

### *Reaction Pathway Calculations*

We identified the reactant and product of $H_2$ dissociation in a 1×1×2 supercell of [Na(taci)][$e^-$] by relaxing molecular and dissociated $H_2$ at the PBE-D3(BJ) level of theory. These calculations relaxed only the atomic positions at fixed cell volume. We calculated an initial guess for the reaction pathway with the Nudged Elastic Band (NEB) method[28] using a relatively coarse maximum force tolerance of 0.1 eV/Å and an SCF convergence criterion of $10^{-5}$ eV. This calculation used nine total images including the reactant and product states. From this calculation, we used the dimer method[29] as implemented in VTST Tools to identify the true transition state. An intrinsic reaction coordinate (IRC) calculation[30] was used to confirm and map out the minimum energy pathway. The dimer method and IRC calculations were run with a maximum force tolerance of 0.02 eV/Å. Partial atomic charges were calculated via the Bader partitioning method[31] for the initial, transition, and final states. Since the dissociation of an isolated $H_2$ molecule does not have a well-defined transition state, single-point calculations were run at fixed H–H bond distances for $H_2$ dissociation in vacuum and $[Na(taci)]^+$. For the deprotonation calculation, a climbing image NEB calculation[32] was carried out with a maximum force tolerance of 0.02 eV/Å.

## Supplementary Figures and Tables

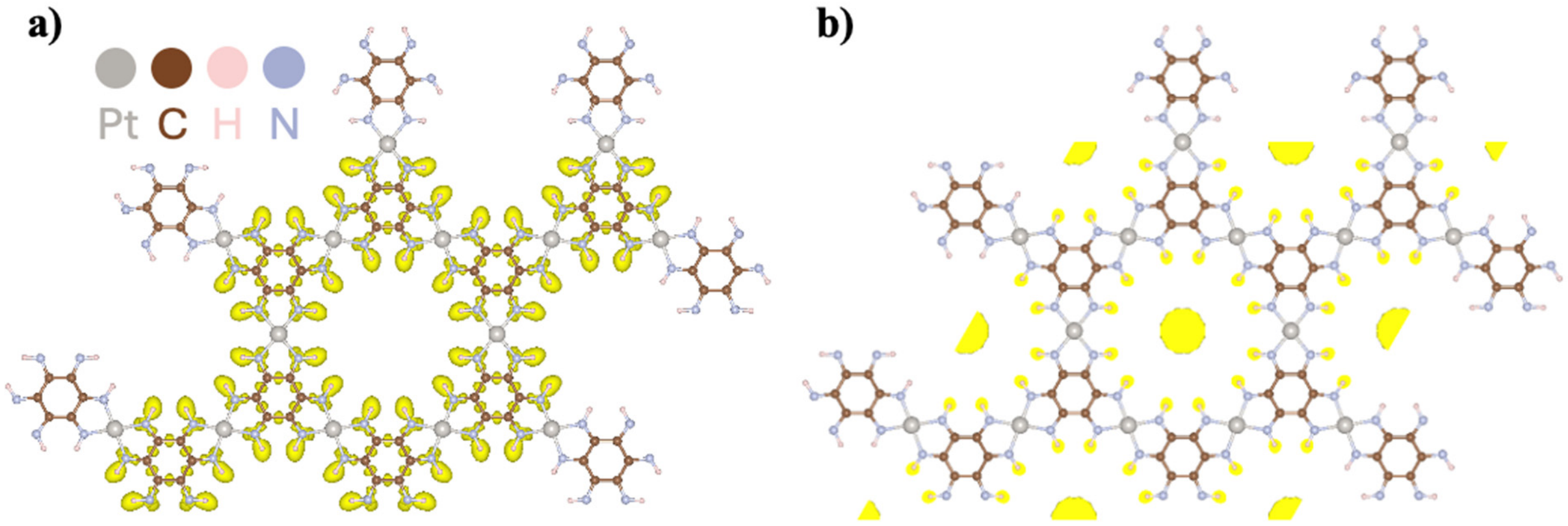

**Figure S1**. Visualization of the ELF of $Pt_3(hib)_2$ (a) as a neutral system (ELF isosurface = 0.771) and (b) doped with one additional electron per unit cell (ELF isosurface = 0.956). Only the ELF spin-down channel is visualized in panel b, as the electride state is not present in the spin-up channel.

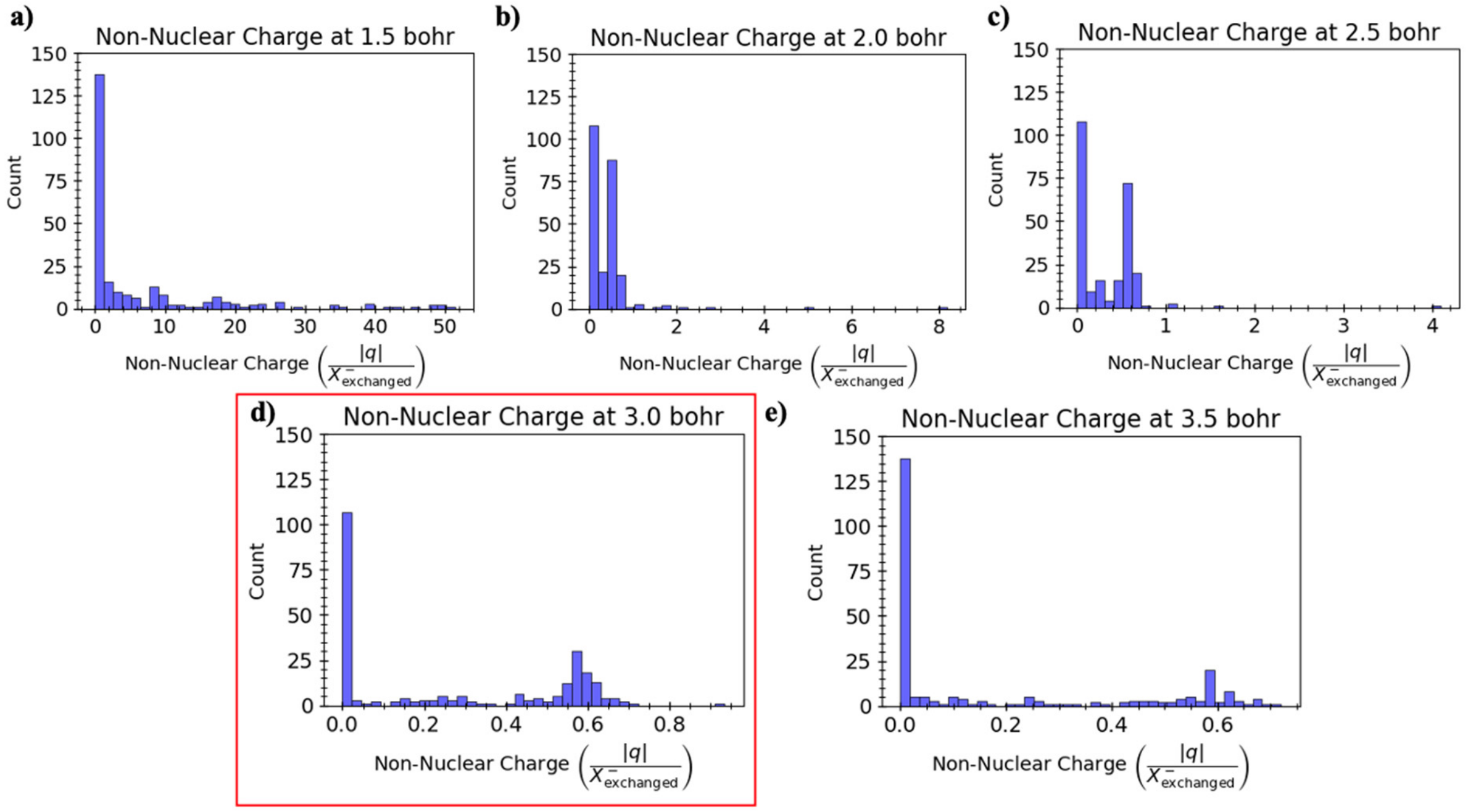

**Figure S2**. Distribution of non-nuclear charge across different cutoff values for ELF basins to be considered non-nuclear. The following distances were considered: (a) 1.5 bohr, (b) 2.0 bohr, (c) 2.5 bohr, (d) 3.0 bohr, and (e) 3.5 bohr. A value of 3.0 bohr was chosen (red outline) as a suitable cutoff value because it eliminated the false-positive detection of non-nuclear ELF basins upon visual inspection. Lower values result in non-physical non-nuclear charges. All data was obtained at the PBE-D3(BJ) level of theory.

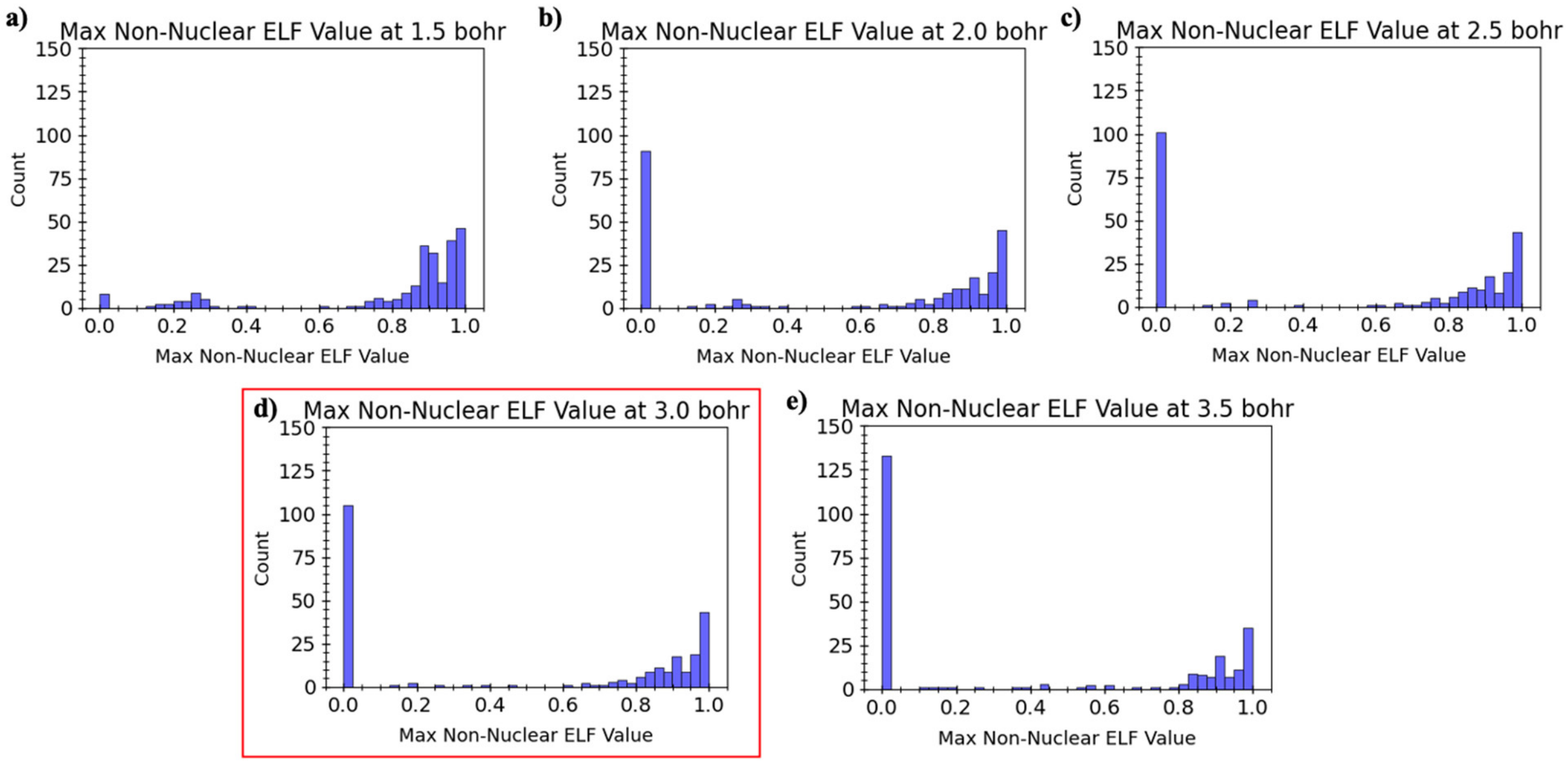

**Figure S3.** Distribution of maximum non-nuclear ELF value across different cutoff values for ELF basins to be considered non-nuclear. The following distances were considered: (a) 1.5 bohr, (b) 2.0 bohr, (c) 2.5 bohr, (d) 3.0 bohr, and (e) 3.5 bohr. A value of 3.0 bohr was chosen (red outline) as a suitable cutoff value in conjunction with integrated charge density data. All data was obtained at the PBE-D3(BJ) level of theory.

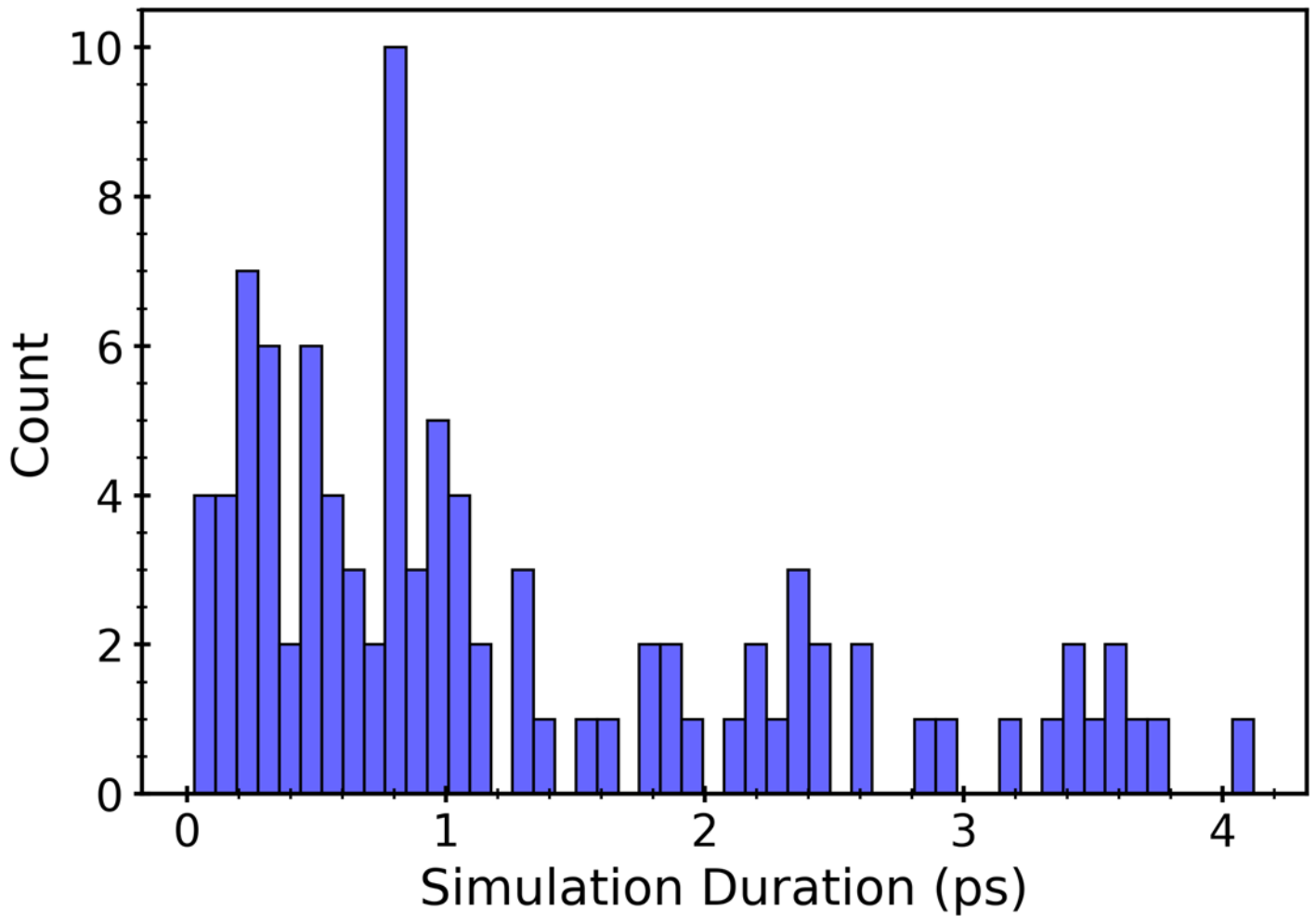

**Figure S4.** Distribution of AIMD (*NVT*, 300 K) simulation durations for the 92 distinct MOF electride candidates. Energies and forces were evaluated at the PBE-D3(BJ) level, and a timestep of 0.5 fs was used. The mean simulation length was 1.26 ps.

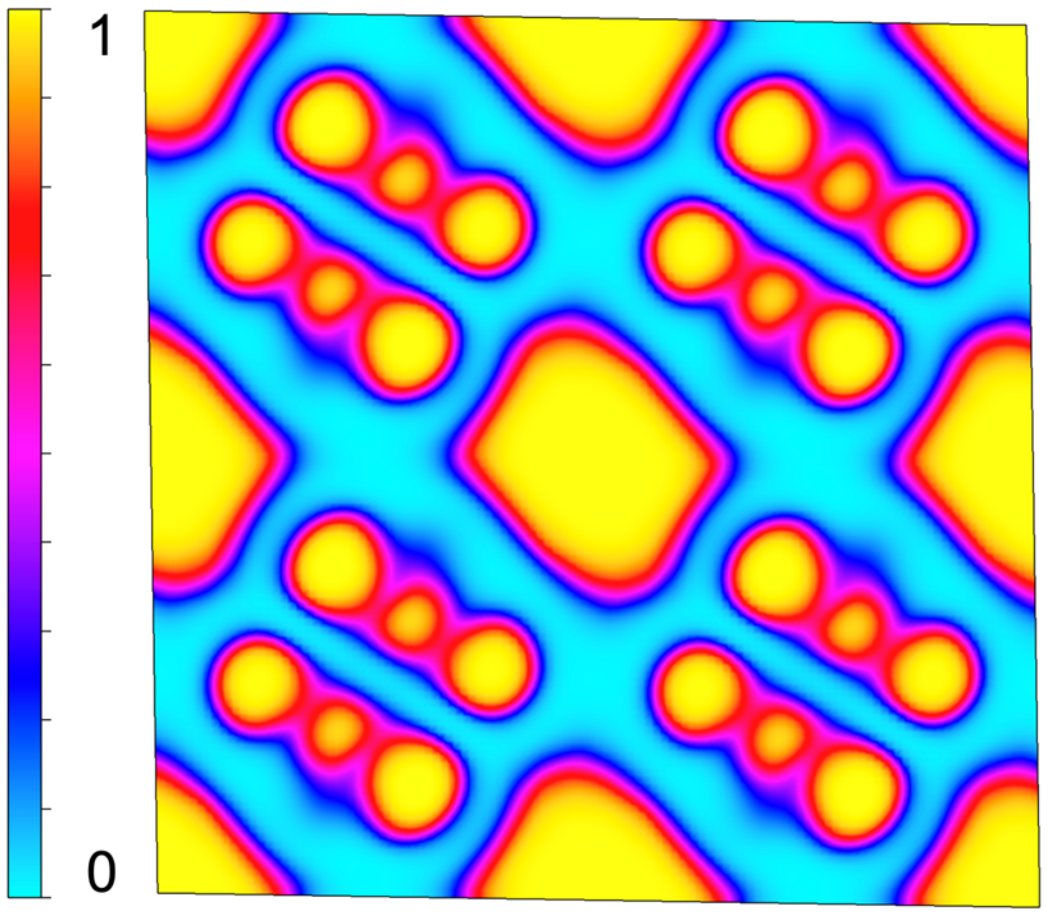

**Figure S5.** Visualizations of the ELF for [Na(1,4-dioxane)$_3$][e$^-$] ((0,0,1) plane – 0 Å from origin, spin-down channel). Electride state seen in yellow at the center of the unit cell. Electronic structure information in this figure was calculated at the HSE06 level of theory.

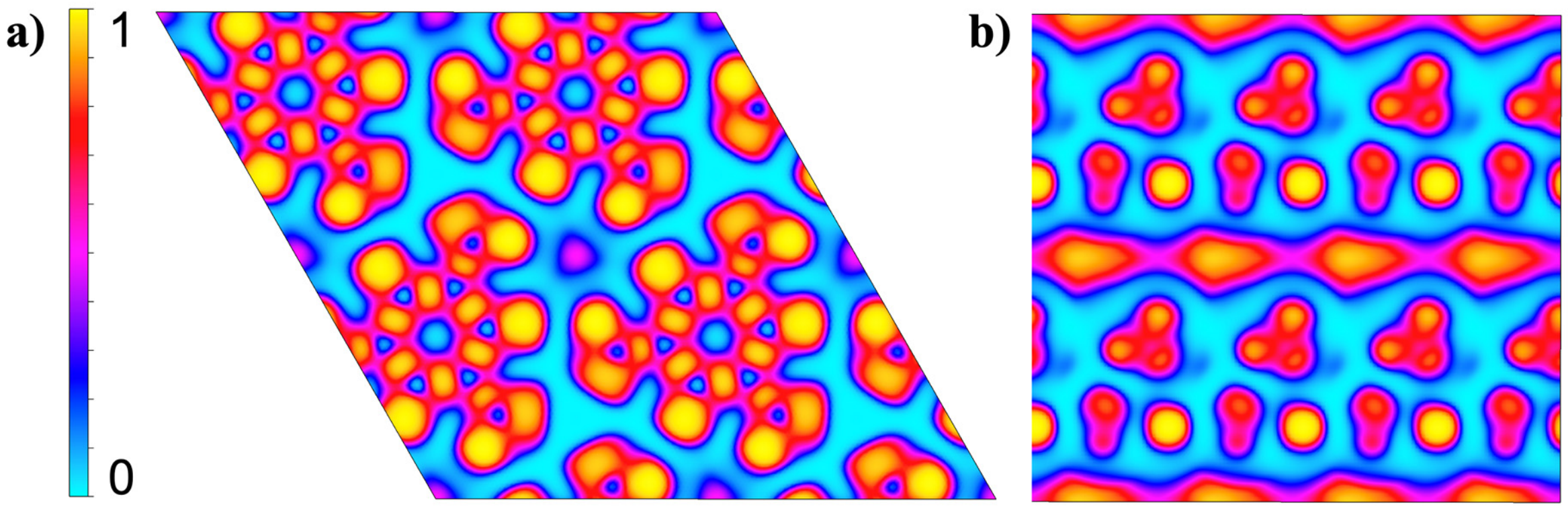

**Figure S6.** Visualizations of the ELF for [Na(taci)][e⁻] (a) ((0,0,1) plane – 0 Å from origin, spin-up channel). (b) ((0,1,0) plane – 0 Å from origin, spin-up channel). Electride state seen in pink/yellow at the center of the unit cell. Electronic structure information in this figure was calculated at the HSE06 level of theory.

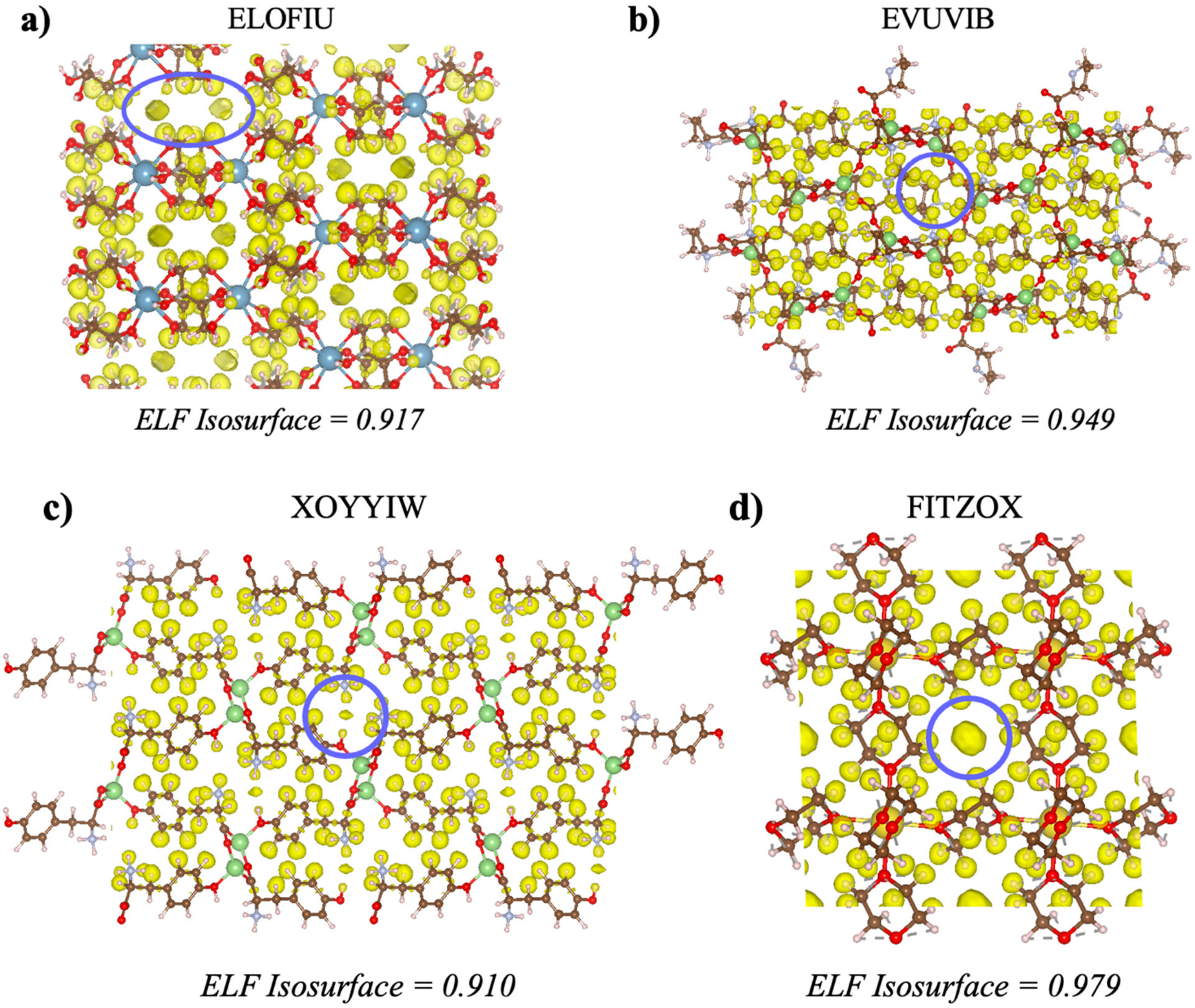

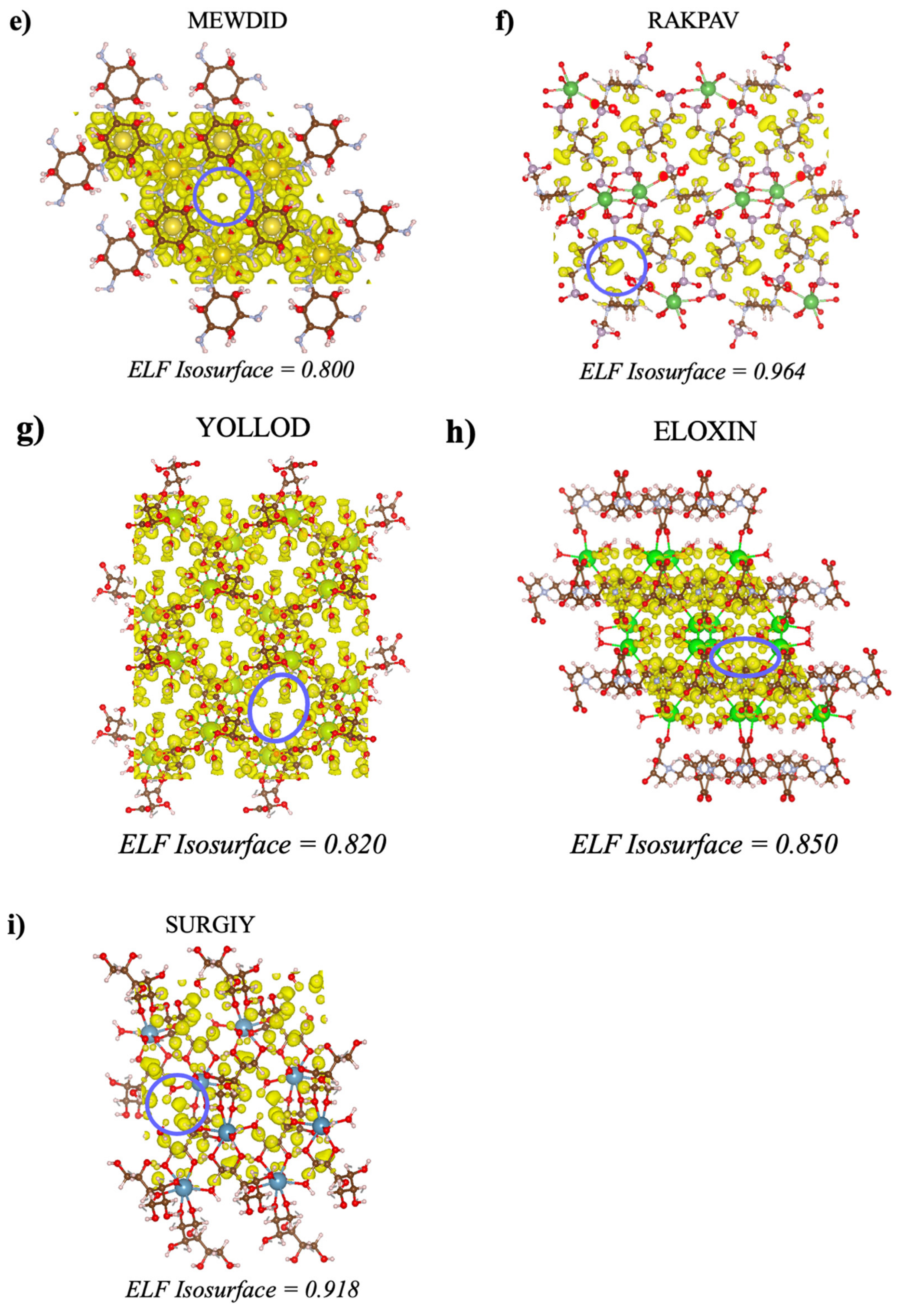

**Figure S7.** Visualizations of the ELF for the nine 3D-connected MOF electride candidates highlighted in this manuscript. Electride states circled in blue. We note that the free electron is sometimes difficult to view within a 2D frame and have included the ELFCAR files on NOMAD for reference. All electronic structure information in this figure is calculated at the PBE-D3(BJ) level of theory.

**Table S1.** Non-nuclear charge and maximum non-nuclear ELF value for the 67 MOF electride candidates (as well as other relevant data). The maximum ELF value is reported based on consideration of both the spin-up and spin-down channels. The non-nuclear charge is normalized on a per-anion-exchanged basis. All data is at the PBE-D3(BJ) level of theory. The complete data in machine-readable form is available on Figshare, as described in the main text.

| Refcode | Non-Nuclear ELF Max | Non-Nuclear Charge (\|q\|/ $X^{-}_{exchanged}$) | Formula | Dim. | Na(15-crown-5) $\Delta E_{rxn}$ (eV/$X^{-}_{exchanged}$) | Na(naphthalenide) $\Delta E_{rxn}$ (eV/$X^{-}_{exchanged}$) |
|---|---|---|---|---|---|---|
| AFEKEC | 0.988 | 0.288 | Li4 H48 C28 N16 O8 | 1 | -0.27 | -0.69 |
| AZIPOQ | 0.873 | 0.567 | Li4 H44 C24 N12 O12 | 1 | -0.09 | -0.51 |
| AZIPUW | 0.877 | 0.585 | Li4 H44 C24 N12 O12 | 1 | -0.16 | -0.58 |
| AZIQAD | 0.977 | 0.601 | Li2 H22 C12 N6 O6 | 1 | -0.04 | -0.45 |
| BOPDAM | 0.931 | 0.581 | Ba4 H84 C36 N12 O24 | 1 | -0.05 | -0.47 |
| CALCLA02 | 0.92 | 0.502 | Ca2 H22 C4 O12 | 1 | -0.20 | -0.62 |
| DEKSEU | 0.749 | 0.604 | Ca1 H22 C12 N6 O6 | 1 | -0.22 | -0.64 |
| DEKYAW | 0.836 | 0.583 | Ca2 H44 C24 N12 O12 | 1 | 0.01 | -0.40 |
| DEKYEA | 0.737 | 0.596 | Ca1 H22 C12 N6 O6 | 1 | -0.28 | -0.70 |
| ELOFEQ | 0.933 | 0.301 | Ca2 H36 C8 O16 | 1 | -0.37 | -0.78 |
| ELOFIU | 0.983 | 0.159 | Ca2 H40 C16 O16 | 3 | -0.28 | -0.70 |
| ELOXIN | 0.876 | 0.246 | Sr4 H48 C20 N4 O26 | 3 | -0.33 | -0.74 |
| EVUVIB | 0.972 | 0.468 | Li4 H72 C40 N8 O16 | 3 | -0.23 | -0.64 |
| EVUWAU | 0.87 | 0.527 | Li8 H112 C48 N16 O32 | 2 | -0.21 | -0.63 |
| EZOMOY | 0.99 | 0.631 | La8 H128 C64 N16 O80 | 2 | -0.14 | -0.56 |
| FIPVEI | 0.848 | 0.555 | Ca1 H18 C6 N2 O6 | 1 | 0.00 | -0.42 |
| FIPVOS | 0.976 | 0.583 | Mg1 H18 C6 N2 O6 | 1 | -0.21 | -0.63 |
| FIPWEJ | 0.915 | 0.582 | Ca4 H84 C36 N12 O24 | 1 | -0.05 | -0.47 |
| FIPXIO | 0.862 | 0.555 | Mg4 H100 C36 N12 O32 | 1 | -0.08 | -0.50 |
| FITZOX | 0.999 | 0.686 | Na1 H24 C12 O6 | 3 | 0.25 | -0.16 |
| FRCPCA01 | 0.994 | 0.62 | Ca1 H28 C12 O14 | 1 | -0.20 | -0.62 |
| GEXQAE | 0.831 | 0.427 | Ca2 H48 C24 N8 O12 | 2 | -0.22 | -0.64 |
| HAVPUP | 0.615 | 0.199 | Cu2 H44 C12 N12 | 1 | -0.94 | -1.36 |
| HEFXEV | 0.808 | 0.479 | Li2 H16 C8 N4 O6 | 2 | -0.43 | -0.85 |
| HIPQIG | 0.999 | 0.675 | K1 H26 C10 N2 O6 | 1 | 0.34 | -0.07 |
| HOBWAX | 0.911 | 0.579 | Na4 H60 C24 N4 O12 | 1 | 0.09 | -0.33 |
| IDANUZ | 0.971 | 0.72 | K1 Si8 H64 C36 O34 | 1 | -0.40 | -0.81 |
| ILAYIE | 0.956 | 0.487 | Y4 H64 C28 N40 O28 | 1 | -0.17 | -0.59 |
| IYAQAD | 0.951 | 0.584 | Ca2 H56 C24 O24 | 1 | -0.10 | -0.51 |
| IYAQEH | 0.955 | 0.584 | Ca2 H56 C24 O24 | 1 | 0.02 | -0.40 |
| JAZNEE | 0.998 | 0.67 | Na1 H24 C10 N2 O5 | 2 | 0.09 | -0.33 |
| JUKMEJ | 0.832 | 0.586 | Ca1 H26 C10 N2 O6 | 1 | -0.27 | -0.68 |
| JUKNAG | 0.772 | 0.627 | Ca4 H104 C40 N8 O24 | 1 | -0.17 | -0.58 |
| KEHVOL | 0.994 | 0.622 | Ca1 H28 C12 O14 | 1 | -0.07 | -0.48 |
| LEQVEK | 0.972 | 0.264 | K2 H56 C24 N8 O16 | 1 | 0.11 | -0.31 |
| MEWDID | 0.89 | 0.567 | Na2 H30 C12 N6 O6 | 3 | -0.06 | -0.48 |
| MISJOQ | 0.85 | 0.548 | Li4 H44 C20 N4 O12 | 1 | -0.11 | -0.53 |
| NARZEM | 0.822 | 0.592 | Sr4 H40 C16 S4 N4 O24 | 2 | -0.24 | -0.66 |
| NEPJAU | 0.995 | 0.595 | Li8 H134 C40 N30 O2 | 2 | -0.07 | -0.48 |
| NUNBAC | 0.991 | 0.451 | Ca4 H80 C48 N16 O24 | 1 | -0.14 | -0.56 |
| NUNBEG | 0.821 | 0.509 | Ca2 H36 C12 N4 O14 | 1 | -0.49 | -0.91 |
| OMAPIE | 0.992 | 0.602 | Mg1 H48 C24 N6 O6 | 2 | -0.15 | -0.57 |
| OQEGAU | 0.87 | 0.53 | Ca2 H60 C24 N4 O16 | 1 | -0.29 | -0.71 |
| OROFIM | 0.972 | 0.546 | Ca4 H40 C24 N2 O24 | 1 | -0.31 | -0.73 |
| PANPIH | 0.967 | 0.563 | Ca4 H112 C44 N8 O28 | 1 | -0.35 | -0.77 |
| RAKPAV | 0.992 | 0.679 | La4 P12 H88 C36 N12 O36 | 3 | -0.10 | -0.52 |
| SARCAC02 | 0.916 | 0.58 | Ca4 H84 C36 N12 O24 | 1 | -0.17 | -0.59 |
| SAZPUE | 0.872 | 0.434 | Cd4 H80 C24 I4 N16 | 2 | 0.13 | -0.28 |
| SURGIY | 0.982 | 0.611 | Ca4 H72 C24 O32 | 3 | -0.26 | -0.68 |
| UTEFIL | 0.795 | 0.574 | Li4 H60 C24 N4 O12 | 1 | -0.26 | -0.68 |
| UTEFOR | 0.819 | 0.592 | Li4 H52 C20 N4 O12 | 1 | -0.29 | -0.71 |
| UTEHAF | 0.774 | 0.303 | Li4 H60 C24 N4 O12 | 1 | -0.33 | -0.75 |
| VALCAC01 | 0.777 | 0.626 | Ca4 H104 C40 N8 O24 | 1 | -0.24 | -0.65 |
| VEDCUE | 0.985 | 0.625 | Li4 H56 C24 N8 O16 | 1 | -0.05 | -0.46 |
| VEDDEP | 0.868 | 0.523 | Li4 H56 C24 N8 O16 | 1 | -0.09 | -0.50 |
| VIPSUK | 0.992 | 0.42 | Zn2 H48 C52 N24 Cl2 O4 F8 | 2 | -0.23 | -0.65 |
| VOCGAY | 0.894 | 0.263 | Ca2 H64 C32 N8 O12 | 2 | -0.22 | -0.64 |
| VOCGOM | 0.983 | 0.256 | Mg2 H64 C32 N8 O12 | 2 | -0.23 | -0.64 |
| WACWUW | 0.878 | 0.465 | Zn8 H72 C24 S12 N12 | 2 | -0.31 | -0.73 |
| WIFWEQ | 0.998 | 0.651 | Na8 H144 C48 N16 O32 | 2 | 0.18 | -0.23 |
| WOCGED | 0.986 | 0.674 | Na2 H30 C12 N2 O6 | 1 | -0.17 | -0.58 |
| XELGUS | 0.986 | 0.309 | La8 H128 C32 Cl16 O56 | 1 | -0.21 | -0.63 |
| XETCOS | 0.925 | 0.545 | Ca1 H28 C12 O14 | 1 | 0.21 | -0.20 |
| XOYYIW | 0.95 | 0.439 | Li4 H44 C36 N4 O12 | 3 | -0.40 | -0.81 |
| XOZVOY | 0.991 | 0.42 | Na9 H216 C108 O108 | 1 | -0.25 | -0.67 |
| YOLLOD | 0.879 | 0.566 | La4 H40 C16 O36 | 3 | -0.35 | -0.77 |
| ZZZJTA01 | 0.994 | 0.645 | Sr1 H28 C12 O14 | 1 | -0.18 | -0.59 |

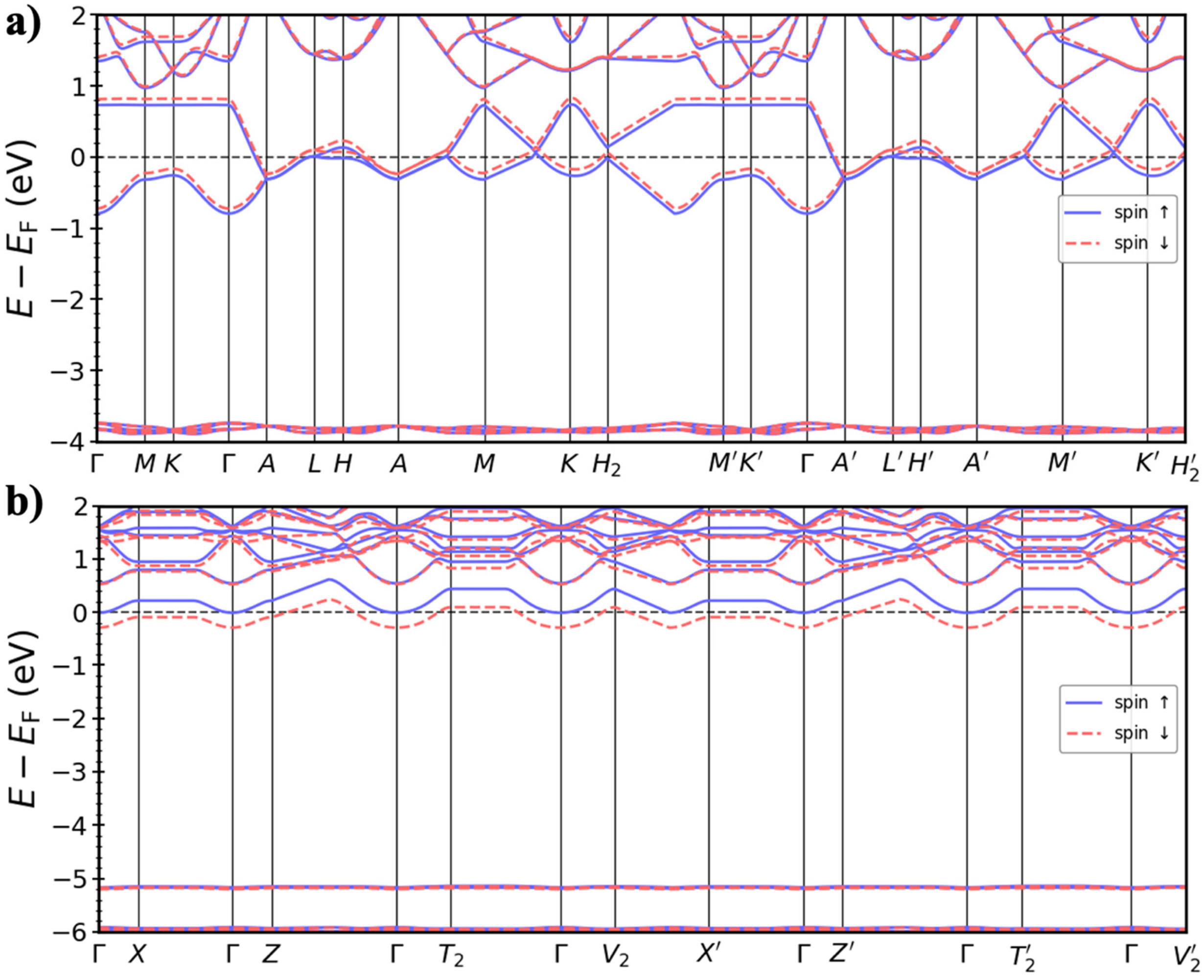

**Figure S8.** Band structure calculations at the PBE-D3(BJ) level of theory for two MOFs with significant electride character. The band(s) corresponding to the pore-confined electron(s) are immediately below or crossing through the Fermi level. (a) [Na(taci)][$e^-$], (b) [Na(1,4-dioxane)$_3$][$e^-$].

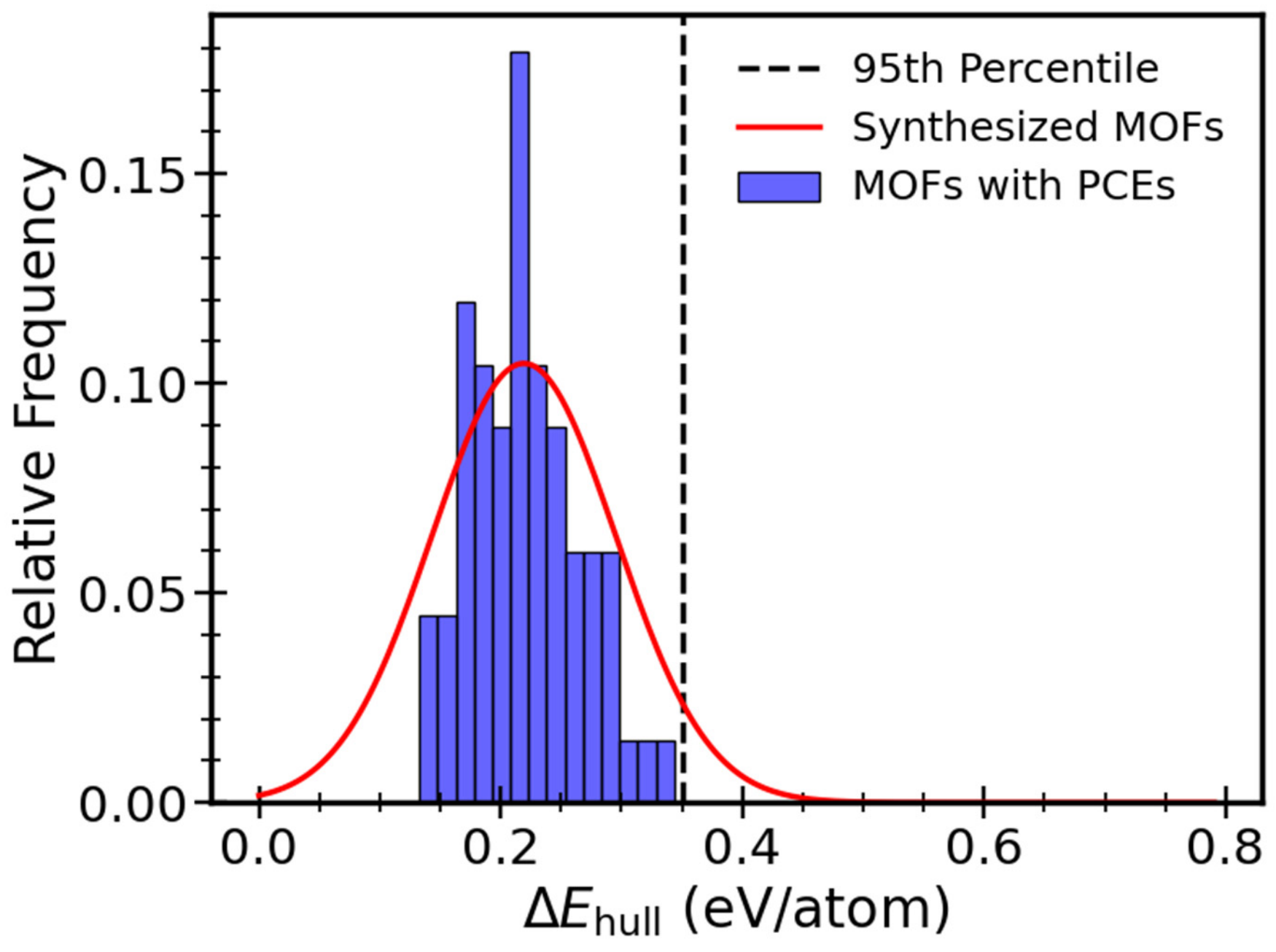

**Figure S9**. $\Delta E_{hull}$ distribution for the 67 MOF electride candidates found via the high-throughput study presented in this work. $\Delta E_{hull}$ values are calculated from structures re-relaxed using the `QMOFSet` settings from QuAcc with a *k*-spacing of 0.4 Å$^{-1}$.[33] The $\Delta E_{hull}$ distribution of synthesized MOFs found in the QMOF database[34,35] is shown in red. A threshold of 0.35 eV/atom was established by Dallmann et al.[33] based on the 95$^{th}$ percentile of synthesized MOFs from the QMOF database, above which $\Delta E_{hull}$ values correspond to MOFs that are unlikely to be synthesizable. The 67 MOF electride candidates highlighted fall within the red distribution under said threshold.

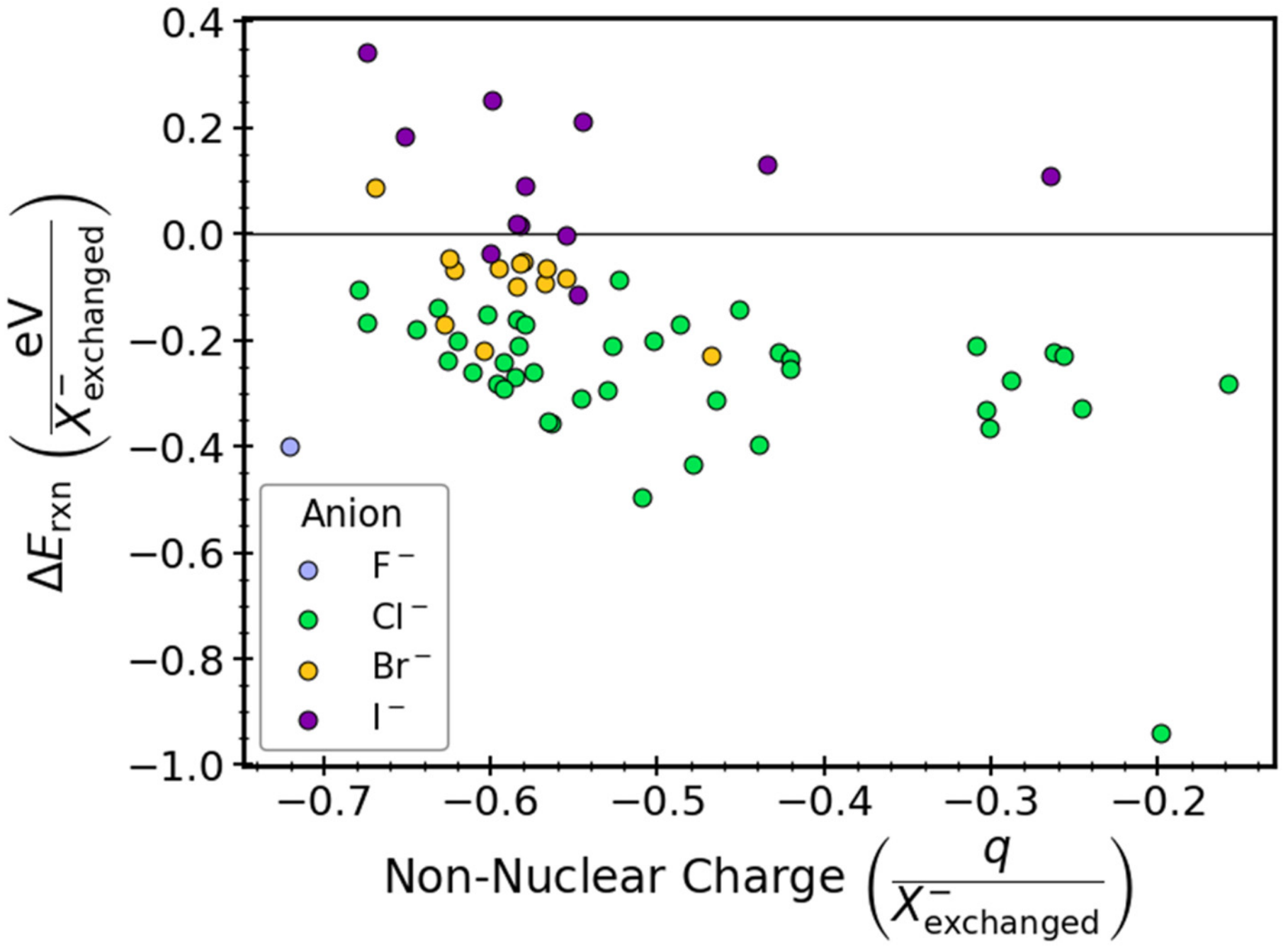

**Figure S10**. $\Delta E_{rxn}$ (using the Na(15-crown-5) scheme) vs. non-nuclear charge across different anions exchanged for the 67 MOFs with significant electride character. The labeled anions represent those in the original crystal structure reported in the Cambridge Structural Database.

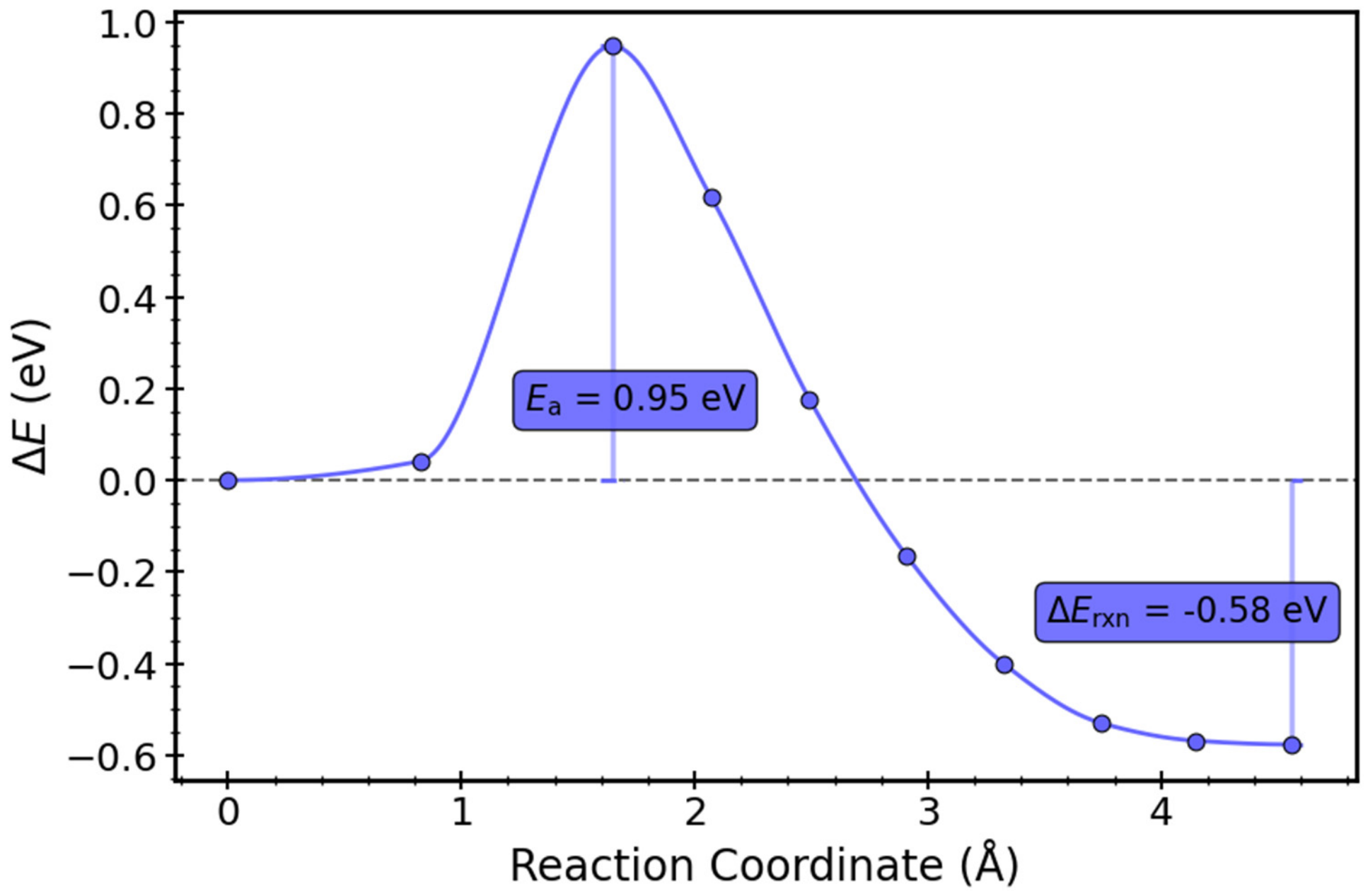

**Figure S11**. Minimum energy pathway from a climbing image nudged elastic band calculation for the deprotonation of an –OH group in [Na(taci)][$e^-$]. Deprotonation of the $–NH_2$ group was found to be thermodynamically and kinetically less favorable.

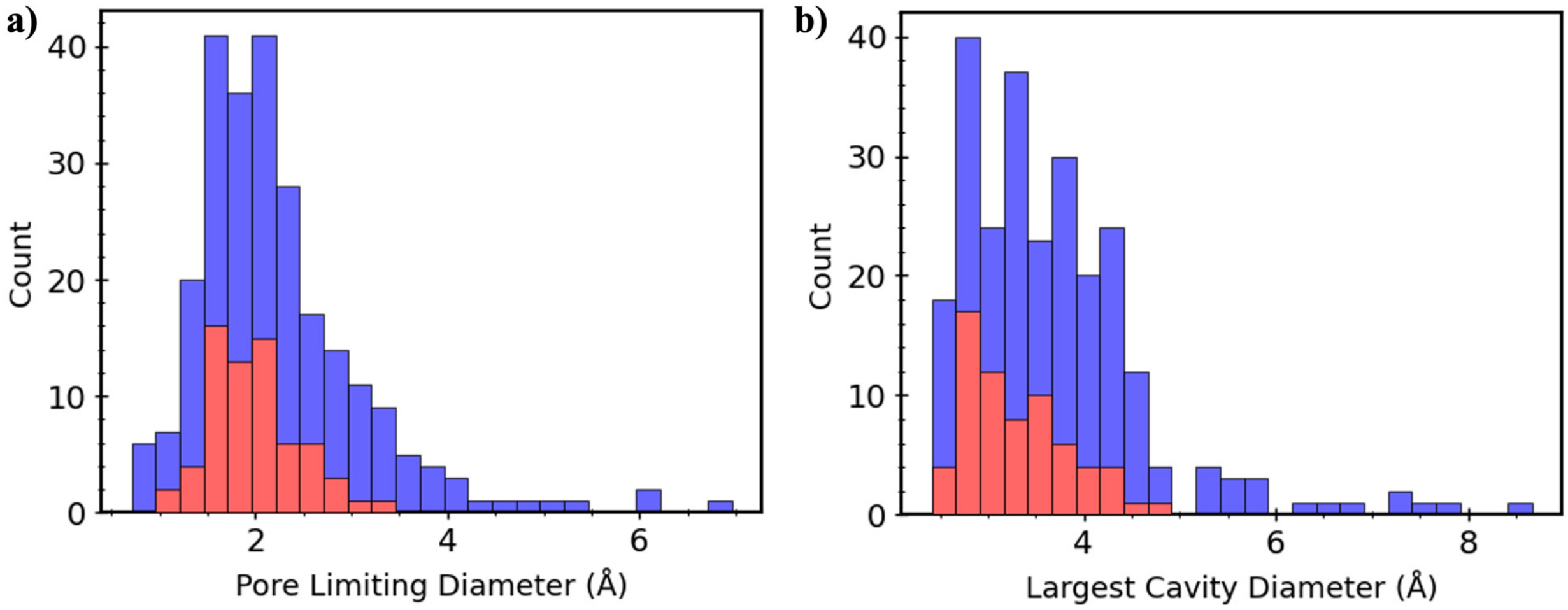

**Figure S12**. Porosity metrics for the 250 MOFs with converged structures after anion–electron exchange (blue) and the 67 MOF electride candidates (red). (a) Pore limiting diameter (PLD) and (b) largest cavity diameter (LCD).

## References

(1) Kresse, G.; Furthmüller, J. Efficient Iterative Schemes for Ab Initio Total-Energy Calculations Using a Plane-Wave Basis Set. *Phys. Rev. B* **1996**, *54* (16), 11169–11186. https://doi.org/10.1103/physrevb.54.11169

(2) Kresse, G.; Furthmüller, J. Efficiency of Ab-Initio Total Energy Calculations for Metals and Semiconductors Using a Plane-Wave Basis Set. *Comput. Mater. Sci.* **1996**, *6* (1), 15–50. https://doi.org/10.1016/0927-0256(96)00008-0

(3) Rosen, A. S.; Goyal, N.; Ayers, B.; Bansal, V.; Baratta, J. H.; Blau, S. M.; Chiang, Y.; Choi, S.; Cohen, O. A.; Dallmann, B.; Demeyere, T.; Engler, W.; Fang, Y.-W.; Furrick, I.; Huerta, E.; Kim, H.; Kondo, H.; Kumar, A.; Kwon, J.; Mamun, O.; Musgrave, C. B.; Oliaei, H.; Saha, A.; Sarpa, D.; Shi, B. X.; Shi, Y.; Wang, X.; Wexler, R. B. Democratizing Atomistic Simulation Workflows for the AI Era with the Quantum Accelerator. **2026**. https://doi.org/10.48550/ARXIV.2609.33823

(4) Larsen, A. H.; Mortensen, J. J.; Blomqvist, J.; Castelli, I. E.; Christensen, R.; Dułak, M.; Friis, J.; Groves, M. N.; Hammer, B.; Hargus, C.; Hermes, E. D.; Jennings, P. C.; Jensen, P. B.; Kermode, J.; Kitchin, J. R.; Kolsbjerg, E. L.; Kubal, J.; Kaasbjerg, K.; Lysgaard, S.; Maronsson, J. B.; Maxson, T.; Olsen, T.; Pastewka, L.; Peterson, A.; Rostgaard, C.; Schiøtz, J.; Schütt, O.; Strange, M.; Thygesen, K. S.; Vegge, T.; Vilhelmsen, L.; Walter, M.; Zeng, Z.; Jacobsen, K. W. The Atomic Simulation Environment—a Python Library for Working with Atoms. *J. Phys. Condens. Matter* **2017**, *29* (27), 273002. https://doi.org/10.1088/1361-648x/aa680e

(5) Rosen, A. S.; Gallant, M.; George, J.; Riebesell, J.; Sahasrabuddhe, H.; Shen, J.-X.; Wen, M.; Evans, M. L.; Petretto, G.; Waroquiers, D.; Rignanese, G.-M.; Persson, K. A.; Jain, A.; Ganose, A. M. Jobflow: Computational Workflows Made Simple. *J. Open Source Softw.* **2024**, *9* (93), 5995. https://doi.org/10.21105/joss.05995

(6) Perdew, J. P.; Burke, K.; Ernzerhof, M. Generalized Gradient Approximation Made Simple. *Phys. Rev. Lett.* **1996**, *77* (18), 3865–3868. https://doi.org/10.1103/PhysRevLett.77.3865

(7) Becke, A. D.; Johnson, E. R. Exchange-Hole Dipole Moment and the Dispersion Interaction. *J. Chem. Phys.* **2005**, *122* (15), 154104. https://doi.org/10.1063/1.1884601

(8) Grimme, S.; Ehrlich, S.; Goerigk, L. Effect of the Damping Function in Dispersion Corrected Density Functional Theory. *J. Comput. Chem.* **2011**, *32* (7), 1456–1465. https://doi.org/10.1002/jcc.21759

(9) Grimme, S.; Antony, J.; Ehrlich, S.; Krieg, H. A Consistent and Accurate Ab Initio Parametrization of Density Functional Dispersion Correction (DFT-D) for the 94 Elements H-Pu. *J. Chem. Phys.* **2010**, *132* (15), 154104. https://doi.org/10.1063/1.3382344

(10) Blöchl, P. E. Projector Augmented-Wave Method. *Phys. Rev. B* **1994**, *50* (24), 17953–17979. https://doi.org/10.1103/PhysRevB.50.17953

(11) Hinuma, Y.; Pizzi, G.; Kumagai, Y.; Oba, F.; Tanaka, I. Band Structure Diagram Paths Based on Crystallography. *Comput. Mater. Sci.* **2017**, *128*, 140–184. https://doi.org/10.1016/j.commatsci.2016.10.015

(12) Togo, A.; Shinohara, K.; Tanaka, I. Spglib: A Software Library for Crystal Symmetry Search. arXiv March 13, 2024. https://doi.org/10.48550/arXiv.1808.01590

(13) Momma, K.; Izumi, F. *VESTA 3* for Three-Dimensional Visualization of Crystal, Volumetric and Morphology Data. *J. Appl. Crystallogr.* **2011**, *44* (6), 1272–1276. https://doi.org/10.1107/S0021889811038970

(14) Evans, D. J.; Holian, B. L. The Nose–Hoover Thermostat. *J. Chem. Phys.* **1985**, *83* (8), 4069–4074. https://doi.org/10.1063/1.449071
(15) Burton, L. A. Best Practices for Modelling Electrides. arXiv 2025. https://doi.org/10.48550/ARXIV.2512.24989
(16) Heyd, J.; Scuseria, G. E.; Ernzerhof, M. Hybrid Functionals Based on a Screened Coulomb Potential. *J. Chem. Phys.* **2003**, *118* (18), 8207–8215. https://doi.org/10.1063/1.1564060
(17) Krukau, A. V.; Vydrov, O. A.; Izmaylov, A. F.; Scuseria, G. E. Influence of the Exchange Screening Parameter on the Performance of Screened Hybrid Functionals. *J. Chem. Phys.* **2006**, *125* (22), 224106. https://doi.org/10.1063/1.2404663
(18) Moghadam, P. Z.; Li, A.; Wiggin, S. B.; Tao, A.; Maloney, A. G. P.; Wood, P. A.; Ward, S. C.; Fairen-Jimenez, D. Development of a Cambridge Structural Database Subset: A Collection of Metal–Organic Frameworks for Past, Present, and Future. *Chem. Mater.* **2017**, *29* (7), 2618–2625. https://doi.org/10.1021/acs.chemmater.7b00441
(19) Bruno, I. J.; Cole, J. C.; Edgington, P. R.; Kessler, M.; Macrae, C. F.; McCabe, P.; Pearson, J.; Taylor, R. New Software for Searching the Cambridge Structural Database and Visualizing Crystal Structures. *Acta Crystallogr. B* **2002**, *58* (3), 389–397. https://doi.org/10.1107/S0108768102003324
(20) Jin, X.; Jablonka, K. M.; Moubarak, E.; Li, Y.; Smit, B. MOFChecker: A Package for Validating and Correcting Metal–Organic Framework (MOF) Structures. *Digit. Discov.* **2025**, *4* (6), 1560–1569. https://doi.org/10.1039/D5DD00109A
(21) Gibaldi, M.; Kapeliukha, A.; White, A.; Woo, T. K. Incorporation of Ligand Charge and Metal Oxidation State Considerations into the Computational Solvent Removal and Activation of Experimental Crystal Structures Preceding Molecular Simulation. *J. Chem. Inf. Model.* **2025**, *65* (1), 275–287. https://doi.org/10.1021/acs.jcim.4c01897
(22) George, J.; Horton, M. K.; Kaplan, A. D.; Kavanagh, S. R.; Riebesell, J.; Rosen, A. S.; Yang, H.; Adams, F.; Amachraa, M.; Andriuc, O.; Ashour, O. A.; Bajaj, S.; Banyas, E.; Barroso-Luque, L.; Berket, K.; Bernstein, N.; Bhardwaj, A.; Bhatti, A. H. U.; Blau, S. M.; Bocklund, B.; Bonkowski, A.; Broberg, D.; Brunin, G.; Bystrom, K.; Cardot, C. A.; Chang, R.; Chen, C.; Chen, W.; Chen, X.; Chiang, Y.; Choudhary, K.; Cohen, O. A.; Costa, M. D.; Dacek, S. T.; de Jong, M.; Demeyere, T.; Dozier, A. K.; Dunn, A.; Dwaraknath, S.; Dylla, M. T.; Eimre, K.; Epstein, A. R.; Ertural, C.; Esters, M.; Evans, M. L.; Falke, J.; Fekete, A.; Field, B.; Fonari, A.; Frey, N. C.; Gaines, D. W.; Ganose, A. M.; Giantomassi, M.; Goodall, R. E. A.; Gostin, P.-F.; Gupta, A.; Ha, V.-A.; Hautier, G.; Heinsdorf, N.; Herbst, M. F.; Hou, T.; Huck, P.; Ikeda, Y.; Jackson, A. J.; Jayaraman, S.; Kearns, F. R.; Kim, S.; Kingsbury, R. S.; Krajewski, A. M.; Kumagai, Y.; Kuner, M. C.; Laderer, W. T.; Li, W.; Li, X.-G.; Liang, Y. H.; Liu, H.; Liu, K.; Liu, M.; Lutfi, A.; Marchand, O.; McDermott, M. J.; Medasani, B. K.; Meng, F.; Miller, K. D.; Miranda, H.; Mohanakrishnan, R. S.; Montoya, J. H.; Moore, G. C.; Morgan, B. J.; Munro, J. M.; Muraoka, K.; Naik, A. A.; O'Leary, W.; Onwuli, A.; Park, J.; Parsons, D. F.; Pei, K.; Petretto, G.; Purcell, T. A. R.; Qi, J.; Qu, X.; Radchenko, I.; Rajput, N. N.; Ricci, F.; Rich, B. S.; Richards, W. D.; Rignanese, G.-M.; Ruan, X.; Vallverdu, G. S.; Schmidt, J.; Shen, J.-X.; Shimogawa, R.; Shinohara, K.; Singh, A. K.; Sivonxay, E.; Smeets, S.; Smidt, T. E.; Sobolev, A.; Spotte-Smith, E. W. C.; Stihl, C.; Sun, W.; Sundberg, J. D.; Suzuki, Y.; Talirz, L.; Tang, H.; Taw, E.; Tezak, C.; Therrien, F.; Thiel, S. D.; Torrisi, S. B.; Turiansky, M. E.; Ueltzen, K.; van Setten, M. J.; Vieten, J.; Vijay, S.; Wagner, N.; Wan, F.; Wang, A.; Wang, J.; Wang, Z.; Ward, L.; Warford, T.; Waroquiers, D.; Wen, M.; Weser, O.; Winner, N.; Winston, D.; Wolloch, M.; Wood, B. M.; Woods-Robinson, R.; Xiao, Y.; Xie,

Y.; Yalcin, F.; Yang, R. X.; Ye, W.; Zarabadi-Poor, P.; Zeng, J.; Zhao, H.; Zheng, K.; Zhu, H.; Zhu, Z.; Zimmermann, N. E. R.; Zügner, D.; Persson, K. A.; Jain, A.; Ong, S. P. Pymatgen: 15 Years of Community Growth, New Functionality, and Future Prospects. **2026**. https://doi.org/10.48550/ARXIV.2610.23538

(23) Otero-de-la-Roza, A.; Johnson, E. R.; Luaña, V. Critic2: A Program for Real-Space Analysis of Quantum Chemical Interactions in Solids. *Comput. Phys. Commun.* **2014**, *185* (3), 1007–1018. https://doi.org/10.1016/j.cpc.2013.10.026

(24) Yu, M.; Trinkle, D. R. Accurate and Efficient Algorithm for Bader Charge Integration. *J. Chem. Phys.* **2011**, *134* (6), 064111. https://doi.org/10.1063/1.3553716

(25) Willems, T. F.; Rycroft, C. H.; Kazi, M.; Meza, J. C.; Haranczyk, M. Algorithms and Tools for High-Throughput Geometry-Based Analysis of Crystalline Porous Materials. *Microporous Mesoporous Mater.* **2012**, *149* (1), 134–141. https://doi.org/10.1016/j.micromeso.2011.08.020

(26) Jain, A.; Ong, S. P.; Hautier, G.; Chen, W.; Richards, W. D.; Dacek, S.; Cholia, S.; Gunter, D.; Skinner, D.; Ceder, G.; Persson, K. A. Commentary: The Materials Project: A Materials Genome Approach to Accelerating Materials Innovation. *APL Mater.* **2013**, *1* (1), 011002. https://doi.org/10.1063/1.4812323

(27) Horton, M. K.; Huck, P.; Yang, R. X.; Munro, J. M.; Dwaraknath, S.; Ganose, A. M.; Kingsbury, R. S.; Wen, M.; Shen, J. X.; Mathis, T. S.; Kaplan, A. D.; Berket, K.; Riebesell, J.; George, J.; Rosen, A. S.; Spotte-Smith, E. W. C.; McDermott, M. J.; Cohen, O. A.; Dunn, A.; Kuner, M. C.; Rignanese, G.-M.; Petretto, G.; Waroquiers, D.; Griffin, S. M.; Neaton, J. B.; Chrzan, D. C.; Asta, M.; Hautier, G.; Cholia, S.; Ceder, G.; Ong, S. P.; Jain, A.; Persson, K. A. Accelerated Data-Driven Materials Science with the Materials Project. *Nat. Mater.* **2025**, 1522–1532. https://doi.org/10.1038/s41563-025-02272-0

(28) Jónsson, H.; Mills, G.; Jacobsen, K. W. Nudged Elastic Band Method for Finding Minimum Energy Paths of Transitions. In *Classical and Quantum Dynamics in Condensed Phase Simulations*; WORLD SCIENTIFIC: LERICI, Villa Marigola, 1998; pp 385–404. https://doi.org/10.1142/9789812839664_0016

(29) Henkelman, G.; Jónsson, H. A Dimer Method for Finding Saddle Points on High Dimensional Potential Surfaces Using Only First Derivatives. *J. Chem. Phys.* **1999**, *111* (15), 7010–7022. https://doi.org/10.1063/1.480097

(30) Fukui, K. The Path of Chemical Reactions - the IRC Approach. *Acc. Chem. Res.* **1981**, *14* (12), 363–368. https://doi.org/10.1021/ar00072a001

(31) Bader, R. F. W.; Tal, Y.; Anderson, S. G.; Nguyen-Dang, T. T. Quantum Topology: Theory of Molecular Structure and Its Change. *Isr. J. Chem.* **1980**, *19* (1–4), 8–29. https://doi.org/10.1002/ijch.198000003

(32) Henkelman, G.; Uberuaga, B. P.; Jónsson, H. A Climbing Image Nudged Elastic Band Method for Finding Saddle Points and Minimum Energy Paths. *J. Chem. Phys.* **2000**, *113* (22), 9901–9904. https://doi.org/10.1063/1.1329672

(33) Dallmann, B.; Saha, A.; Rosen, A. S. Predicting the Thermodynamic Limits of Metal–Organic Framework Metastability. *J. Am. Chem. Soc.* **2026**, *148* (19), 19487–19501. https://doi.org/10.1021/jacs.5c20253

(34) Rosen, A. S.; Fung, V.; Huck, P.; O'Donnell, C. T.; Horton, M. K.; Truhlar, D. G.; Persson, K. A.; Notestein, J. M.; Snurr, R. Q. High-Throughput Predictions of Metal–Organic Framework Electronic Properties: Theoretical Challenges, Graph Neural Networks, and Data

Exploration. *Npj Comput. Mater.* **2022**, *8* (1), 112. https://doi.org/10.1038/s41524-022-00796-6
(35) Rosen, A. S.; Iyer, S. M.; Ray, D.; Yao, Z.; Aspuru-Guzik, A.; Gagliardi, L.; Notestein, J. M.; Snurr, R. Q. Machine Learning the Quantum-Chemical Properties of Metal–Organic Frameworks for Accelerated Materials Discovery. *Matter* **2021**, *4* (5), 1578–1597. https://doi.org/10.1016/j.matt.2021.02.015